\documentclass[a4paper,fleqn,usenatbib]{mnras}

\usepackage[T1]{fontenc}
\usepackage{ae,aecompl}

\usepackage{graphicx}	
\usepackage{amsmath}	
\usepackage{amssymb}	
\usepackage{color}

\usepackage{times}

\definecolor{gray}{rgb}{0.57, 0.64, 0.69}

\title[The mid-plane conditions of discs]{Determining the mid-plane conditions of circumstellar discs using gas and dust modelling: a study of HD~163296}

\author[D.~M.~Boneberg et al.]
{\parbox{\textwidth}{Dominika M. Boneberg$^{1}$\thanks{E-mail: boneberg@ast.cam.ac.uk},
Olja Pani{\'c}$^{1}$\thanks{Royal Society Dorothy Hodgkin Fellow},
Thomas J. Haworth$^{1}$,
Cathie J. Clarke$^{1}$
and Michiel Min$^{2,3}$
}\vspace{0.4cm}\\
\parbox{\textwidth}{
$^{1}$Institute of Astronomy, Madingley Road, Cambridge CB3 0HA, UK\\
$^{2}$SRON Netherlands Institute for Space Research, Sorbonnelaan 2, 3584 CA Utrecht, The Netherlands\\
$^{3}$Astronomical Institute Anton Pannekoek, University of Amsterdam, Science Park 904, 1098 XH Amsterdam, The Netherlands
}}

\date{Accepted XXX. Received YYY; in original form ZZZ}

\pubyear{2016}

\begin{document}
\label{firstpage}
\pagerange{\pageref{firstpage}--\pageref{lastpage}}
\maketitle

\begin{abstract}
The mass of gas in protoplanetary discs is a quantity of great interest for assessing their planet formation potential. Disc gas masses are, however, traditionally inferred from measured dust masses by applying an assumed standard gas-to-dust ratio of $g/d=100$. Furthermore, measuring gas masses based on CO observations has been hindered by the effects of CO freeze-out.
Here we present a novel approach to study the mid-plane gas by combining C$^{18}$O line modelling, CO snowline observations and the spectral energy distribution (SED) and selectively study the inner tens of au where freeze-out is not relevant. We apply the modelling technique to the disc around the Herbig~Ae star HD~163296 with particular focus on the regions within the CO snowline radius, measured to be at 90\,au in this disc.
Our models yield the mass of C$^{18}$O in this inner disc region of $M_{\text{C}^{18}\text{O}}(<90\,\text{au})\sim 2\times10^{-8}$\,M$_\odot$. 
We find that most of our models yield a notably low $g/d<20$, especially in the disc mid-plane ($g/d<1$). Our only models with a more interstellar medium (ISM)-like $g/d$ require C$^{18}$O to be underabundant with respect to the ISM abundances and a significant depletion of sub-micron grains, which is not supported by scattered light observations.
Our technique can be applied to a range of discs and opens up a possibility of measuring gas and dust masses in discs within the CO snowline location without making assumptions about the gas-to-dust ratio.
\end{abstract}

\begin{keywords}
stars: pre-main sequence -- planetary systems: protoplanetary discs -- stars: circumstellar matter -- techniques: interferometric 
\end{keywords}



\section{Introduction}
Protoplanetary discs - discs of gas and dust that surround young pre-main sequence stars - are the birthplaces of planets. With new observational facilities such as the \textit{Atacama Large Millimeter/submillimeter Array} (ALMA) providing data of unprecedented resolution and sensitivity, we have the opportunity to study the disc structure and through it the processes that lead to planet formation in more detail than ever before. However, interpretation of these observations is reliant upon comparison with the
expected emission properties from numerical models of discs. 
Protoplanetary discs consist of gas and dust. In the interstellar medium (ISM), the mass ratio of these two components is canonically assumed to be $g/d$=100 as has been inferred from observations \protect\citep[see e.g.][]{FrerkingEtAl1982,LacyEtAl1994}. 
Due to the lack of observational constraints thereof, this value is often also adopted for discs. \protect\cite{PanicEtAl2008} obtain a range between 25 and 100 for $g/d$ from their modelling of the Herbig Ae/Be star HD~169142, depending on the dust opacity they assume, \protect\cite{MeeusEtAl2010} narrowed down this range to $\sim22-50$. In 51~Oph, \protect\cite{ThiEtAl2013} find a value of $g/d$ consistent with 100, the same holds true for HD~141569 \protect\citep{ThiEtAl2014}. The study by \protect\cite{WilliamsBest2014} of several T Tauri stars yields $g/d$ that are relatively low ($\lesssim 40$; the values they obtain for a few Herbig Ae/Be stars are also rather low, with the exception of HD~163296, where they obtain $g/d=170$).

The gas governs the disc dynamics and motion of the dust, whereas dust provides the opacity to capture the stellar flux, re-radiate it and heat the disc.
In order to understand the structure of discs, it is therefore crucial to study the spatial distribution and properties of both components \protect\citep[see e.g.][]{BeckwithSargent1987,DutreyEtAl1994,IsellaEtAl2007, PanicEtAl2008, QiEtAl2011,PanicEtAl2014}.
Planets are believed to form in the disc mid-plane and thus understanding the disc conditions in these regions is particularly important for constraining models of planet formation \protect\citep[see e.g.][]{BoleyDurisen2010,ForganRice2013}.

Many discs have bright emission in the millimetre (mm) continuum \protect\citep[e.g.][]{BeckwithEtAl1990, DutreyEtAl1996,ManningsSargent1997,AndrewsWilliams2007, AndrewsEtAl2009, AndrewsEtAl2010, QiEtAl2011}, tracing the dust in the disc. Due to the uncertainties associated with $g/d$ and the dust grain properties, inferring the disc mass from continuum measurements is however only a rough approximation. Therefore, molecular emission lines are used alternatively or additionally and allow the inference of spatial and temperature structure.
The most abundant molecule in discs is cold H$_2$ gas, however this is difficult to observe due to the lack of a dipole moment and its low transition probability. Thus, molecules such as  $^{12}$CO, $^{13}$CO and C$^{18}$O and their respective transitions are employed instead.

Abundances of molecular tracers are influenced by the conditions in the disc: for example CO can be photodissociated in the disc atmosphere or frozen out in the disc below a temperature of $T\approx$19\,K \protect\citep{QiEtAl2011}. Recently it has been claimed that the exact value of the freeze-out temperature can vary from disc to disc (e.g. \protect\cite{QiEtAl2015} model the snowline at a temperature of 17\,K in TW Hya and 25\,K in HD~163296) and depends on the chemical history of the ice \protect\citep{GarrodPauly2011}.
Moreover, the transitions of the CO emission lines become optically thick at different heights within the disc, depending on the abundance of the particular isotopologue \protect\citep[e.g.][]{ZadelhoffEtAl2001,DartoisEtAl2003,MiotelloEtAl2014} and thus
optical depth effects compromise the ability to obtain disc masses from
the more abundant species. 
C$^{18}$O is an important diagnostic of the unfrozen part of the disc mass, being much less abundant than other CO species ([$^{16}$O]/[$^{18}$O]=$557\pm30$, \protect\citealt{Wilson1999}). Its transitions in the mm wavelength regime are mostly optically thin throughout the whole disc and thus provide an excellent probe of the disc mid-plane. This is evidently of great importance since most of the gas mass resides
near the disc mid-plane and it is here that planets are expected to form.
However, only a handful of observations of C$^{18}$O exist so far, including AB~Aurigae \protect\citep{SemenovEtAl2005}, HD~169142 \protect\citep{PanicEtAl2008}, MWC480 \protect\citep{AkiyamaEtAl2013}, HD~142527 \protect\citep{PerezEtAl2015} and HD~163296 \protect\citep{QiEtAl2011,RosenfeldEtAl2013}. There are also C$^{18}$O data available on several T Tauri stars studied by \protect\cite{WilliamsBest2014}. Furthermore, there exist observations of C$^{18}$O in TW Hya, V4046 Sgr, DM Tau, GG Tau and IM Lup \protect\citep[see][and references therein]{WilliamsBest2014}.

The \textit{CO snowline radius} is the location in the disc mid-plane at which CO condenses from the gas phase and freezes out on to dust grains. This radius can be observed as a steep decline in the C$^{18}$O density or by the presence of other molecular tracers such as N$_2$H$^+$ and DCO$^+$ (\protect\citealt{QiEtAl2011, MathewsEtAl2013,QiEtAl2013chemistry, QiEtAl2015}; Carney et al., in prep.). However, the formation path of DCO$^{+}$ is not fully understood and this molecule does not probe the disc mid-plane but the entire surface of the 19-21\,K isotherms. Hence \protect\cite{QiEtAl2015} find that DCO$^{+}$ is not a reliable tracer of the CO snowline location. They employ ALMA N$_2$H$^{+}$J=3-2 observations instead that originate predominantly from the mid-plane and are therefore more reliable. 
The presence of gas-phase CO slows down the formation of N$_2$H$^{+}$ and accelerates its destruction. Thus, gas-phase N$_2$H$^{+}$ exists in regions where CO is depleted, so the N$_2$H$^{+}$ emission will be distributed in a ring whose inner radius marks the CO snowline location. Therefore,  \protect\cite{QiEtAl2015} propose that observations of C$^{18}$O and N$_2$H$^{+}$ are very powerful as they directly probe the temperature of the disc mid-plane. 
This is important for calculations of the vertical hydrostatic equilibrium in discs which crucially depend on the conditions in this disc region.
However, the exact freeze-out temperature of CO is not known unambiguously, depends on the conditions of the environment and is assumed to be between $\sim17\,$K \protect\citep[][in TW Hya]{QiEtAl2013} and $\sim 30\,$K \protect\citep[][in an embedded protostar]{JorgensenEtAl2015}. Also, the composition of the ice will influence the freeze-out temperature ($\sim 20\,$K for pure CO ice, $\sim 30\,$K for mixed CO-H$_2$O ice, \protect\cite{CollingsEtAl2004}). 
In addition, the gas pressure can also have an impact on the freeze-out temperature \protect\citep{FraySchmitt2009}; however Stammler et al. (in prep.) find that changes in the gas pressure in the disc mid-plane are not sufficient to shift the CO snowline radius by amounts which would cause an observable effect in our observations.
Nevertheless, measurements of the snowline location are important as they give constraints on the mid-plane temperature profiles of discs.

Another important tool for studying protoplanetary discs is the \textit{spectral energy distribution} (SED) that combines independent measurements in a range of wavelength regimes that trace different parts of the disc \protect\citep[see e.g.][ Pani{\'c} et al. (subm.) for studies of the influence of disc parameters on the resulting SED]{BossYorke1996, Dullemond2002, Meijer2008}.
As the dust content of the disc influences its opacity and thus determines how much stellar flux can be intercepted and re-radiated by the disc, the SED crucially depends on the properties and vertical distribution of dust.
Thus a combination of C$^{18}$O observations, additional data on the CO snowline radius and the SED provide a powerful combination of observables to model protoplanetary discs, combining independent measurements of both gas and dust.

In this paper we model the disc around the $2.3\,\text{M}_\odot$ \protect\citep{QiEtAl2011} Herbig Ae star HD~163296, that is assumed to have an age of $\sim5$Myr \protect\citep{NattaEtAl2004}.
It is situated at a distance of about $d=122$\,pc \protect\citep{VanDenAnckerEtAl1998} with a luminosity of $L=37.7\,\text{L}_\odot$ and an effective temperature of $T_{\text{eff}}=9250$\,K \protect\citep{TillingEtAl2012}.
We list the observational properties of both the star and disc in Table \ref{tab:params_disc_star}.
\begin{table}
	\begin{center}
		\bgroup
		\def\arraystretch{1.}
		\begin{tabular}{l|c}
			\hline \bf{Stellar properties} & \bf{Value} \\ 
			\hline
			\hline Spectral type$^{2}$  &  A1\\ 
			\hline Mass$^2$ $M_{*}$ & 2.3\,M$_\odot$  \\ 
			\hline Effective temperature$^1$ $T_{\text{eff}}$ & 9250\,K  \\ 
			\hline Luminosity$^1$ $L_{*}$ & 37.7\,L$_\odot$ \\  
			\hline Distance$^2$ $d$ & 122\,pc \\
			\hline Age$^3$ $t$ & $\sim$5\,Myr\\
			\hline 
			\hline \bf{Disc parameters} & \bf{Value} \\ 
			\hline
			\hline Outer radius (scattered light)$^4$ $R_\text{out, sc}$ &  $\sim500$\,au\\  
			\hline Outer radius (continuum, 850\,$\mu$m)$^{5,6}$ $R_\text{out, cont}$ & $\sim240-290$\,au  \\ 
			\hline Outer radius (CO observations)$^5$ $R_\text{out, CO}$ & $\sim550$\,au\\
			\hline CO snowline radius$^{7}$ $R_\text{sl}$ & $90$\,au\\
			\hline
		\end{tabular} 
		\egroup
		\caption {Observational stellar and disc properties of HD~163296 from: $^1$\protect\cite{TillingEtAl2012}, $^{2}$\protect\cite{QiEtAl2011}, $^{3}$\protect\cite{NattaEtAl2004}, $^4$\protect\cite{GradyEtAl2000}, $^5$\protect\cite{deGregorioMonsalvoEtAl2013}, $^6$\protect\cite{GuidiEtAl2016} and $^7$\protect\cite{QiEtAl2015} We use a Kurucz model for the star. }
		\label{tab:params_disc_star} 
	\end{center}
\end{table}
Interestingly, the outer radius of the disc as inferred from CO emission studies   \protect\citep{QiEtAl2011,deGregorioMonsalvoEtAl2013} and scattered light \protect\citep{GradyEtAl2000} is about double the value of the disc outer radius observed in the continuum \protect\citep{deGregorioMonsalvoEtAl2013}.
It is worth noting that HD~163296 is a relatively bright Herbig Ae star ($L_{*}$=37.7\,L$_\odot$, \protect\citet{TillingEtAl2012}), thus, its disc is comparatively warm and its C$^{18}$O line emission strong. Furthermore, the disc is observed to have a gap in polarized light at $R_\text{gap}\sim 70$\,au \protect\citep{GarufiEtAl2014}.
Its molecular lines (mostly CO) and continuum have been studied in detail in the mm and sub-mm \protect\citep{ManningsSargent1997,NattaEtAl2004,IsellaEtAl2007,QiEtAl2011} and recently also with ALMA \protect\citep{deGregorioMonsalvoEtAl2013, RosenfeldEtAl2013,MathewsEtAl2013,FlahertyEtAl2015,QiEtAl2015,GuidiEtAl2016}. 

\protect\cite{RosenfeldEtAl2013, deGregorioMonsalvoEtAl2013} and \protect\cite{QiEtAl2015} employed ALMA data for modelling of the disc of HD~163296 but we are the first to use the C$^{18}$O J=2-1 data to model disc parameters.
\protect\cite{RosenfeldEtAl2013} focused mainly on modelling CO and $^{13}$CO, whereas \protect\cite{deGregorioMonsalvoEtAl2013} analysed the Band 7 data ($^{12}$CO J=3--2 and continuum) and \protect\cite{GuidiEtAl2016} were most interested in the dust properties and hence the continuum observations. \protect\cite{QiEtAl2015} also studied the C$^{18}$O emission, but they were mostly interested in analysing the snowline location and comparing the C$^{18}$O and N$_2$H$^{+}$ emission. In addition, \protect\cite{FlahertyEtAl2015} used the available C$^{18}$O data, but focused on the turbulence in the disc.
We describe the relevant \textsc{ALMA} observations in the next section and stress that we base our modelling
on the available ALMA C$^{18}$O data as a crucial ingredient.

\protect\cite{QiEtAl2011} had inferred a snowline radius $R_\text{sl}\sim155$\,au from $^{13}$CO observations which was consistently also derived by \protect\cite{MathewsEtAl2013} from DCO$^{+}$ observations. However, more recent studies by \protect\cite{QiEtAl2015} find a snowline radius $R_\text{sl}\sim90$\,au from both N$_2$H$^{+}$ and C$^{18}$O ALMA observations. 

We do not aim to provide one best-fitting model for the disc around HD~163296, but rather want to emphasize the degeneracies in the parameters of the modelling process and propose a way to overcome them. Our main goal is to investigate the mid-plane gas temperature and density in this disc using a novel modelling approach.
In Section~\ref{sec:observations}, we summarize and discuss the observations. In Section~\ref{sec:discussion}, we describe in detail our modelling process and all the steps involved. In Section~\ref{subsec:models}, we specify the models we obtain, their implications and potential degeneracies and also discuss their properties. We summarize our findings and conclusions in Section~\ref{sec:summaryconclusions}.
\section{\textsc{ALMA} observations}
\label{sec:observations}
\subsection{Description of observations}
Science verification data of HD~163296 were taken by ALMA in Band~6 and 7 \protect\citep{RosenfeldEtAl2013}. The ALMA observations are provided as 3D fits cubes with two spatial and one spectral axis (velocity/frequency) on the ALMA Portal\footnote{\texttt{https://almascience.nrao.edu/alma-data/science-\linebreak verification/overview}}. 
There are calibrated and cleaned data with a resolution of $\sim0.7\,''$ ($\sim$85\,au at $d=122$\,pc) available for $^{12}$CO J=2--1, $^{13}$CO J=2--1 and C$^{18}$O J=2--1 (all Band~6), as well as for $^{12}$CO J=3--2 (Band~7). We list the corresponding RMS noise values and beam sizes in Table~\ref{tab:resolrms}.
 We use the \textit{Common Astronomy Software Applications} \textsc{CASA} software package version 4.4.0 \protect\citep{McMullinEtAl2007}, to analyse the respective transitions. Amongst the transitions listed above, the C$^{18}$O J=2-1 is relatively unexplored and the one on which our work focuses.
 We employ the already self-calibrated and cleaned Science Verification data provided on the ALMA Portal.
\begin{table}
	\begin{center}
		\bgroup
		\def\arraystretch{1.}
		\begin{tabular}{l|c|c}
			\hline \bf{Molecular lines} & \bf{Synthesized beam}[$''$] &  \bf{rms ($\sigma$)}[Jy\,beam$^{-1}$] \\ 
			\hline
			\hline C$^{18}$O J=2--1 (SV) & $0.73 \times 0.58$ & $ 2\times10^{-2}$ \\ 
			\hline $^{12}$CO J=2--1 (SV) & $0.68 \times 0.55$ & $ 5\times10^{-2}$ \\ 
			\hline $^{13}$CO J=2--1 (SV) & $0.72 \times 0.57$ & $ 3\times10^{-2}$ \\ 
			\hline $^{12}$CO J=3--2 (SV)& $0.65 \times 0.42$ & $ 5\times10^{-2}$ \\  
			\hline 
		\end{tabular} 
		\egroup
		\caption {Summary of the available ALMA observations (molecular lines in Bands~6 and 7)}
		\label{tab:resolrms} 
	\end{center}
\end{table}

C$^{18}$O J=2-1 observations (Band 6) of HD~163296 (RA$=17^h56^m21\mbox{\ensuremath{.\!\!^{s}}}281$, Dec.$=-21^\circ 57'22\mbox{\ensuremath{.\!\!^{\prime\prime}}}36$; J2000) were taken with 24 ALMA antennas (12\,m) in 2012, on June 9 and 23 and July 7 with baselines spanning 20--400\,m and a total on-source time of 84\,min \protect\citep{RosenfeldEtAl2013}. For a detailed summary of the spectral windows and calibrations, see \protect\cite{RosenfeldEtAl2013}. The beam size of the reduced and cleaned C$^{18}$O data is  $0.73'' \times 0.58''$, the spectral resolution is $0.33$\,km\,s$^{-1}$ ($\sim0.24$\,MHz) with 150 channels, ranging from 219.571\,GHz to 219.534\,GHz, where the rest frequency of the transition is 219.56\,GHz. 
We plot the integrated emission and intensity weighted velocity maps of C$^{18}$O J=2-1 in Figure~\ref{fig:mom0maps} and describe them in more detail in the next section. We find an integrated intensity of C$^{18}$O J=2-1 of $6.2\pm0.4$\,Jy\,km\,s$^{-1}$, which is consistent with the values obtained by \protect\cite{QiEtAl2011,QiEtAl2015} and \protect\cite{RosenfeldEtAl2013}.

\subsection{Spatial structure of the emission}
\begin{figure*}
	\centering{
		\includegraphics[width=\textwidth]{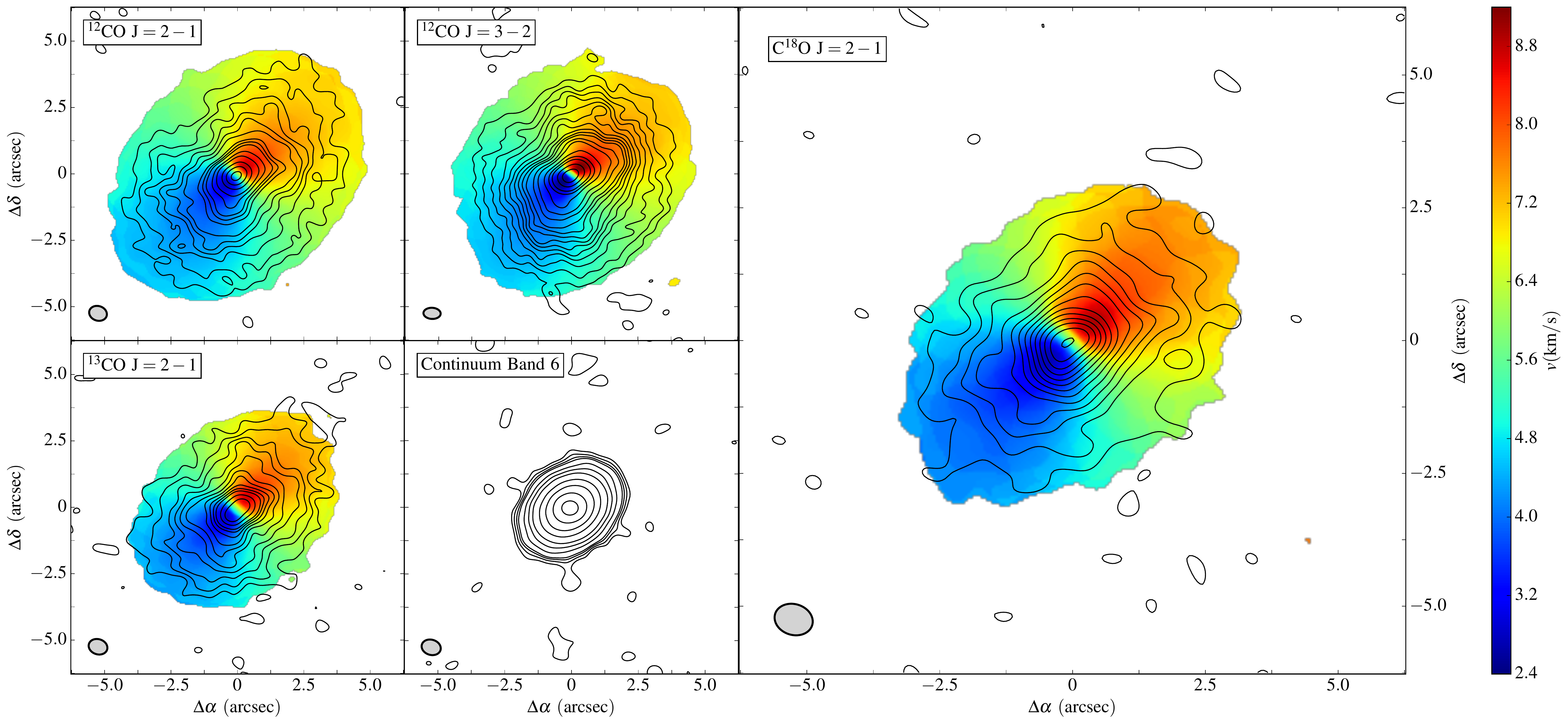}
		\caption{Integrated line emission (contours) and intensity-weighted velocity (colour) maps of $^{12}$CO J=2--1, $^{12}$CO J=3--2, $^{13}$CO J=2--1, C$^{18}$O J=2--1 and continuum map of Band 6. The contours are levels of $(2,4,6,8,...)\times\sigma$ noise. The innermost contour has the following levels: $\sim100\times\sigma$ ($^{12}$CO J=2--1), $\sim100\times\sigma$  ($^{12}$CO J=3--2), $\sim28\times\sigma$ ($^{13}$CO J=2--1), $\sim52\times\sigma$ (C$^{18}$O J=2--1) and $\sim796\times\sigma$ (continuum). The velocity maps discard the data at a level $\lesssim5\times\sigma$ noise. The respective $\sigma$ are given in Table~\ref{tab:resolrms}. The synthesized beam is plotted in the bottom left corner of each panel.}
		\label{fig:mom0maps}
	}
\end{figure*}
We plot the frequency-integrated intensity maps
and intensity-weighted velocity maps for both transitions of $^{12}$CO, as well as C$^{18}$O and $^{13}$CO in Figure \ref{fig:mom0maps}\footnote{As already pointed out in \protect\cite{RosenfeldEtAl2013}, the SV data of $^{12}$CO $\mbox{J=3--2}$ have a velocity offset and are falsely centred around a velocity of 6.99\,km\,s$^{-1}$ instead of the systemic velocity of 5.8\,km\,s$^{-1}$. We have taken this offset into account for the respective velocity map.}.
We will focus in more detail on C$^{18}$O in this paper, but show all of these maps here as we explore the disc geometry also by using these other molecular species. Additionally, we present the continuum map of Band 6.
From the extent of the disc in the panels of Figure \ref{fig:mom0maps}, it is clear that the molecular species and the continuum trace different parts of the disc. The CO isotopes of different abundances trace down to varying depths in the disc, due to their different opacities.

Using the \textsc{CASA} software package, we can determine the position angle (PA) of the disc from C$^{18}$O observations (image deconvolved from beam), which we find to be PA=(132.8$\pm$3.4)\,$^\circ$. This is in agreement with what other studies have found from CO and continuum observations \protect\citep{QiEtAl2011, RosenfeldEtAl2013}.
Fitting a 2D Gaussian to the spatial profile of the emission with \textsc{CASA} allows us to determine the inclination of the disc for the different tracers shown in Figure \ref{fig:mom0maps} (where $i=0$ corresponds to the disc being face-on). From the C$^{18}$O emission, we obtain an inclination of $i=$(47.9$\pm$1.6)\,$^\circ$, which is comparable to the inclination of $i=$44$^{\circ}$ used by \protect\cite{QiEtAl2011}.
For the other molecular species, we performed the same analysis and obtained the values given in Table~\ref{tab:PAincl}. The value we find for the PA from the $^{12}$CO J=3--2 is comparable with the one from \protect\cite{deGregorioMonsalvoEtAl2013}, however, we find a larger inclination in comparison to their value (38\,$^{\circ}$). 
The PA and inclination for the continuum emission in Band 6 and 7 are slightly lower than the values obtained from the gas lines. However, gas and dust can trace regions of the disc with different outer radii. It might thus be possible that the inner regions of the disc have a different inclination. Also, the calculations of the PA and inclination from the CO emission might be influenced by the fact that the line emission seems to have a slightly boxy shape in comparison to the ellipses in the continuum (see Figure~\ref{fig:mom0maps}). For our models, we will adapt a PA=132\,$^{\circ}$ and an inclination of $i$=48\,$^{\circ}$, which is widely in agreement with the values obtained from the fits.
\begin{table}
	\begin{center}
		\bgroup
		\def\arraystretch{1.}
		\begin{tabular}{l|c|c}
			\hline \bf{Molecular species and continuum} & \bf{PA[$^\circ$]} & \bf{Inclination} $i$[$^\circ$]  \\ 
			\hline
			\hline $^{12}$CO J=2--1  & 138.0$\pm$2.0  & 48.4$\pm$2.3\\ 
			\hline $^{13}$CO J=2--1 & 133.7$\pm$2.7 & 46.5$\pm$1.5\\ 
			\hline C$^{18}$O J=2--1 &  132.8$\pm$3.4 & 47.9$\pm$1.6\\ 
			\hline Continuum Band 6 & 131.4$\pm$2.1 & 42.8$\pm$0.1\\ 
			\hline $^{12}$CO J=3--2& 140.4$\pm$1.9 & 44.7$\pm$0.9\\  
			\hline Continuum Band 7 & 130.3$\pm$1.1 & 43.1$\pm$0.1\\
			\hline 
		\end{tabular} 
		\egroup
		\caption {Molecular species and continuum emission in Bands~6 and 7 and the respective PAs and inclinations ($i=0$ is face-on) including their errors obtained from their integrated intensity maps with \textsc{CASA}}
		\label{tab:PAincl} 
	\end{center}
\end{table}
\section{Modelling}
\label{sec:discussion}
\subsection{Physical models}
\label{subsec:modelling}
\subsubsection{Modelling the 2D structure of the disc}
\label{subsec:2dstructuremcmax}
Our modelling process is two-fold: we first model the 2D temperature and density structure of the disc using the radiative transfer code \textsc{MCMax} \protect\citep{MinEtAl2009}, ensuring that our models match the observed SED and CO snowline radius. We then take these models as an input to the \textsc{torus} code \protect\citep{Harries2000}  which performs molecular line radiative transfer; we use synthetic C$^{18}$O line profiles to further narrow down the range of viable models. 
	We primarily aim to determine the magnitude of various parameter degeneracies, rather than to calculate a single best-fit model.
\subparagraph{\bf{\textsc{MCMax}}}
We use the 3D radiative transfer code \textsc{MCMax}, which self-consistently calculates a 2D temperature and density structure of the model with Monte Carlo radiative transfer \protect\citep{MinEtAl2009}. 
The input parameters are the radial variation of the gas and dust surface
density (fixed as being proportional to $r^{-p}$ where
$p$ is in the range $1-1.2$), the total dust mass (and grain
size distribution, $a_\text{min}$ and $a_\text{max}$ ), the gas-to-dust ratio, $g/d$ and the turbulent mixing parameter $\alpha_\text{turb}$.
We then use MCMax to iteratively compute the temperature, and from it the resulting vertical profile of the gas density.
This profile satisfies  hydrostatic equilibrium normal to the disc plane in the gas
and thermal equilibrium in the dust (assuming the gas and dust temperatures are equal).
The vertical profile of the dust (including dust settling) is obtained by solving a diffusion
equation for each dust particle size bin
(see below) and normalizing the vertically averaged value
of the gas-to-dust ratio at each radius to the input value of $g/d$.

The solution is self-consistent in the sense that the dust and gas profiles are
not independently prescribed: the dust affects the hydrostatic
equilibrium of the gas by setting the temperature whereas the gas profile
affects the degree of dust settling and hence, through variation of the
amount of starlight intercepted, the temperature profile of the dust.
For each iteration on the thermal structure of the dust, photon packages emitted from the star (which is the source of heating) are followed through the disc. They are (re-)absorbed, re-emitted and scattered off the dust grains multiple times. This is the primary source of heating for the dense regions of the disc which are of interest for the C$^{18}$O emission. 
This treatment would probably not be suitable for the inner few au of the mid-plane, where viscous heating usually dominates. The mass accretion rate of HD~163296 is derived to be within the range (0.8-4.5) $\times 10^{-7}$\,M$_\odot$yr$^{-1}$  \protect\citep{GarciaLopezEtAl2015} from Br-$\gamma$ observations, so depending on its value, viscous heating could potentially be important in the very inner disc regions ($\sim$few au). Given that we are interested in mid-plane regions further out, we do not take this effect into account.

Based on the stellar properties and the disc structure, a 2D temperature profile for the disc is thus obtained in thermal equilibrium.
\textsc{MCMax} then iterates the gas density profile so as to obtain vertical hydrostatic equilibrium given by:
\begin{equation}
\frac{\mbox{d}P}{\mbox{d}z} = -\rho(r,z)\hspace{1pt} \frac{\mbox{d}F_\text{grav, z}}{\mbox{d}z} \hspace{2pt},
\end{equation}
where $P$ is the pressure, $\rho$ the gas density and $F_\text{grav, z}$ the gravitational potential in $z$-direction. 
Dust settling is included in \textsc{MCMax} by solving a diffusion
equation for each grain size as detailed in \protect\cite{MuldersDominik2012}. We explore values of the turbulent mixing parameter in \textsc{MCMax} between $\alpha_\text{turb}=10^{-4}$ and $10^{-2}$, which is a frequently adopted range of values for protoplanetary discs \protect\citep{MuldersDominik2012}. The value of  $\alpha_\text{turb}$ is hard to derive from observations and is assumed to be in the range of $\sim0.5-10^{-4}$ \protect\citep{IsellaEtAl2009}. For HD~163296, \protect\cite{FlahertyEtAl2015} find a value of $\alpha_\text{turb}\leq9\times10^{-4}$ in the upper layers of the outer disc.
In general, this parameter determines the strength of the mixing of the gas and dust components for a given gas-to-dust mass ratio $g/d$. Furthermore, $\alpha_\text{turb}$ is in general lower at low altitudes in the disc \protect\citep{SimonEtAl2015}. Increasing the turbulent mixing parameter leads to a stronger mixing of gas and dust, enabling more small dust grains to be stirred up to the disc atmosphere where they can intercept more stellar light.
All \textsc{MCMax} models with the same $M_\text{gas}\times\alpha_\text{turb}=\text{const.}$ yield exactly the same SED and CO snowline location. This can be understood following the discussion in \protect\cite{YoudinLithwick2007}:
 for a regime where the dimensionless stopping time $\tau_\text{s}=\Omega_\text{k}\times t_\text{stop}$ (with the Keplerian orbital frequency $\Omega_\text{k}$ and the particle stopping time $t_\text{stop}$) is smaller than the dimensionless eddy turnover time $\tau_\text{e}=\Omega_\text{k}\times t_\text{eddy}$ (with $t_\text{eddy}$ the eddy turnover time), i.e. $\tau_\text{s}<\tau_\text{e}<1$, the scale height of particles $H_\text{p}$ divided by the scale-height of the gas $H_\text{gas}$ is given by
 \begin{equation}
\frac{H_\text{p}}{H_\text{gas}} \propto \sqrt{\frac{\alpha_\text{turb}}{\tau_\text{s}}} \hspace{2pt}.
\label{eq:Hp_Hgas}
 \end{equation}
Given that $\tau_\text{s} \propto t_\text{stop}$, Equation~\ref{eq:Hp_Hgas} can be modified in the Epstein regime, where $t_\text{stop} \propto \rho_\text{grain} \times s \times c_\text{s}^{-1} \times \rho_\text{gas}^{-1}$ (with $\rho_\text{grain}$ being the internal grain density, $s$ the grain size, $c_\text{s}$ the sound speed and $\rho_\text{gas}$ the gas density), hence
 \begin{equation}
 \frac{H_\text{p}}{H_\text{gas}} \propto \sqrt{\alpha_\text{turb}\times \rho_\text{gas}} \hspace{2pt}.
 \label{eq:Hp_Hgas2}
 \end{equation}
 This implies that $H_\text{p} \times H_\text{gas}^{-1}$  and thus the temperature and dust structure are kept invariant when $\rho_\text{gas} \times \alpha_\text{turb}$ (and thus $g/d \times \alpha_\text{turb}$) are kept constant.
Consequently, models that fulfil this criterion have exactly the same SED, temperature structure and dust density structure.
 For our modelling process we first run models with a turbulent mixing strength of $\alpha_\text{turb}=10^{-4}$ and fit these to the observed SED, but then run additional calculations for these models, exploring larger values of $\alpha_\text{turb}$ ($10^{-3}$ and $10^{-2}$) while decreasing the gas masses in these models by factors of 10 and 100, accordingly, to keep the temperature structure and SED the same. The new models are named A-E/10 and A-E/100, respectively. We will call all models that have the same dust parameters and constant $\alpha_\text{turb} \times g/d$ models of the same series.

We perform the above iterations using $5\times10^{7}$ photon packages and 350 grid cells in the azimuthal direction and 400 in radial direction. We have checked for convergence in the Monte Carlo radiative transfer calculation by increasing the number of photon packets.
However, since the Poisson noise scales with the square root of
the number of photon packets, this is inefficient. We therefore
stack the average density and temperature structure over the
last few well converged iterations of the \textsc{MCMax} calculation
to reduce both the noise and computational expense. 
\subparagraph{\bf{Modelling the SED}}
In general, each of our models is unique in some aspect (see \textit{distinctive feature} in last column of Table~\ref{tab:3models} where we list the individual model parameters). We make an extreme assumption for one of the varied parameters at a time and then search for the SED fit in order to obtain the wide range of properties without the necessity of doing a complete parameter space exploration.
We generate a range of  models by varying the following parameters: the mass of dust $M_\text{dust}$, the minimum dust grain size $a_\text{min}$, maximum grain size $a_\text{max}$ and the gas-to-dust ratio $g/d$ (alone, as well as in combination with the turbulent mixing strength $\alpha_\text{turb}$). 
We assume in every case that the grains follow a power-law size distribution of
$n(a) \propto a^{-k}$, where $a$ is the grain size. The value for ISM grains, that is often also adopted for discs, is $k=3.5$ (see e.g.\ \protect\citealt{MathisEtAl1977,ClaytonEtAl2003}), which results in most of the dust mass being in the largest grains, while the opacity is provided by the smallest dust.

We choose to vary these five disc parameters in our modelling as they have the biggest effect on the SED \protect\citep[][Pani{\'c} et al., subm.]{Meijer2008}.	
	We explore a range of $M_\text{dust}$ going from the lowest possible value that can still reproduce the mm flux in the SED as we will describe in Section~\ref{subsec:SEDandSnowline} up to $\sim 0.1\%$ of the stellar mass.
We initially vary $g/d$ between $10$ and $200$ to explore an extreme range around the ISM value (while fixing $\alpha_\text{turb}=10^{-4}$). Once we find a combination of $\alpha_\text{turb}$ and $g/d$ that provide a match, we then explore other combinations of these two parameters, taking into account the above described degeneracy of $\alpha_\text{turb} \times g/d$. The grains sizes we assume range from pristine dust (sub-micron-sized) to mm or even cm-sized grains in some cases.

The stellar properties that we use for all of our models are listed in Table~\ref{tab:params_disc_star} and are kept fixed. We use a Kurucz model for the star, which sets the stellar emission. Given the values for the outer radius as inferred from CO observations, we use a value of $R_\text{out}=540$\,au for our models.
We then investigate how our parameter choices affect the resulting SED and the predicted radius of the CO snowline. Rather than finding the single model
that provides the best fit to these observables, we instead identify a range of models that provide an acceptable fit and then, as detailed in the following section, further isolate the models that additionally match the line fluxes in C$^{18}$O.

\subsubsection{Modelling of the C$^{18}$O line emission}
\label{subsubsec:torusAlma}
The models we obtain from the analysis described above are then taken as an input density and temperature structure for the next modelling step.
We use the radiation transport and hydrodynamics code \textsc{torus} to
perform molecular line transfer calculations in this paper \protect\citep[see e.g.][]{Harries2000, RundleEtAl2010, HaworthEtAl2012}. \textsc{torus} is capable of molecular statistical equilibrium calculations and the production of synthetic data cubes (e.g. for one specific molecular transition).
Full details of the main molecular line transfer algorithm are given by \protect\cite{RundleEtAl2010}; we summarize key and new features below. 

We map the gas density and temperature distributions from the 2D spherical \textsc{MCMax} calculations on to the 2D cylindrical \textsc{torus} grid using a bilinear interpolation in $r$ and $\theta$. 
We assume that the gas and dust are thermally coupled (allowing us to map the dust temperature directly on to the gas). 
Although \textsc{torus} is capable of non-local thermodynamic equilibrium (LTE) molecular line transport, for application to these disc models the densities are sufficiently high and the assumption of LTE produces identical results as non-LTE calculations.\footnote{The assumption of LTE is prudent in the mid-plane for mm lines (in which we are interested as they preferentially probe the disc mid-plane), but might not be sufficient for the infrared (IR) lines.}
Therefore, the level populations can be characterized as a Boltzmann distribution, that is as a simple function of temperature 

\begin{equation} 
	\frac{n_i}{\sum_i n_i} = \frac{g_i \exp\left(\frac{-E_i}{kT_{\text{gas}}}\right)}{z\left(T_{\text{gas}}\right)} \hspace{2pt},
	\label{LTEpops}
\end{equation}
where $g_i$ is the statistical weight and $z\left(T_{\text{gas}}\right)={\sum_i g_i \exp\left(\frac{-E_i}{kT_{\text{gas}}}\right)}$ the partition function.
With the level populations computed, synthetic data cubes are calculated using ray tracing \protect\citep{RundleEtAl2010}. 
\textsc{torus} allows for flexible choice of observer viewing angle and spectral/spatial resolution.
We implement a model for freeze-out, whereby the C$^{18}$O abundance drops to a negligible value if the temperature is below the freeze-out temperature (which is $T_\text{mid-plane}$(90\,au)) of the respective model. We list these temperatures in Table~\ref{tab:3models}.
To evaluate the effect of photodissociation of CO by the stellar irradiation, we implement a simple criterion, qualitatively similar to that of \protect\cite{WilliamsBest2014}. We assume that for a CO particle column density of $N_\text{CO} \approx10^{18}$\,cm$^{-2}$ in the line of sight from the star all CO molecules will be photodissociated. This is only a crude estimate, however it allows us to check how much of the total gas mass is affected and to gauge the impact on our model.
The value we adopt implies a larger role for photodissociation than that employed by \protect\cite{WilliamsBest2014} (who use $N_{\text{H}_2}\approx 1.3 \times 10^{21}$\,cm$^{-2}$, corresponding to N$_\text{CO} \approx 1.3 \times 10^{17}$\,cm$^{-2}$ for $f_\text{CO} \approx 10^{-4}$).
We also explore the effect of adopting even larger column density thresholds of $N_\text{CO}$$\approx10^{19}$\,cm$^{-2}$ and $\approx10^{20}$\,cm$^{-2}$, the latter of which certainly exaggerates the effect of photodissociation.

Turbulence affects the line emission to a much lesser extent than the temperature and density do, and these effects are only marginally discerned in observations of higher signal to noise lines, such as those of $^{12}$CO and at a high spectral resolution. Our assumption of turbulent velocity $v_\text{turb}$ therefore does not affect our fit to the C$^{18}$O data. 
The maximum turbulent line broadening possible is set by the sound speed in the outer mid-plane
\begin{equation}
c_\textrm{s}=\sqrt{\frac{k_\text{B} T_\text{mid}(r_\text{out})}{\mu m_\text{H}}} \hspace{2pt},
\label{eq:cs}
\end{equation}
where $k_\text{B}$ is the Boltzmann constant, $T_\text{mid}$ the mid-plane temperature, $\mu$ the mean molecular weight and $m_\text{H}$ the atomic mass of hydrogen.
Following \protect\cite{SimonEtAl2015}, we employ a value of $0.01 - 0.1 c_{\textrm{s}}$, suitable in the outer disc mid-plane, for the turbulent line broadening in \textsc{torus}.
Given that the temperature in the outer mid-plane is approximately $T\approx8$\,K, we find 
\begin{equation}
v_\text{turb} = (0.01 - 0.1) c_{\textrm{s}} \text{(8K)}= (0.0017 - 0.017)\,\text{km s}^{-1} \hspace{2pt},
\label{eq:vturb}
\end{equation}
where $\mu =2.37$. 
The recent study by \protect\cite{FlahertyEtAl2015} also suggests that turbulence is relatively weak in the HD~163296 disc ($v_\text{turb} < 0.03 c_{\textrm{s}}$ in the upper layers of the outer disc), supporting our low value of $v_\text{turb}$. 
Changing $v_\text{turb}$ by a factor of 10 in our models does not alter the fit to the observations of C$^{18}$O, as the data quality does not allow us to probe  $v_\text{turb}$ sufficiently well. Also, since C$^{18}$O is mainly optically thin, turbulent broadening will cause slight smearing of the line profile, but will not affect the line flux. This would be different if we were studying a more optically thick transition like for example CO J=3-2 \protect\citep[see e.g.][]{FlahertyEtAl2015}.
The turbulent line broadening is related to the turbulent mixing parameter $\alpha_\text{turb}$ by
\begin{equation}
v_\text{turb}\sim\sqrt{\alpha_\text{turb}}c_\text{s} \hspace{2pt}.
\end{equation}
The above means that $\alpha_\text{turb}= 10^{-4} - 10^{-2}$ is implicit in this calculation. \protect{\cite{FlahertyEtAl2015}} find $\alpha_\text{turb}< 9.6 \times 10^{-4}$ from their modelling of HD~163296. We explore this range of values of $\alpha_\text{turb}$ in Section~\ref{subsec:SEDandSnowline}, but we hereby stress that for a wide range of $\alpha_\text{turb}$ and corresponding values of $v_\text{turb}$ our fit to the line emission remains unaffected.

Another aspect to take into account is that the fractional abundance (by number density) of C$^{18}$O is uncertain. This abundance is given by
\begin{equation}
f_{\text{C}^{18}\text{O}}=\frac{[\mbox{CO}]}{[\mbox{H}_2]} \times \frac{[\mbox{C}^{18}\mbox{O}]}{[\mbox{CO}]} \hspace{2pt},
\label{eq:uncertc18oab}
\end{equation}
where we assume that $[\mbox{C}^{18}\mbox{O}]/[\mbox{CO}]$= [$^{18}$O]/[$^{16}$O].
Therefore uncertainty in the isotopic ratio of $^{16}$O to $^{18}$O as well as in the fractional abundance of CO have to be taken into account. The abundance of CO is altered due to freeze-out (mid-plane) and photodissociation (surface) (see e.g.\ \protect\citealt{PanicEtAl2008,MiotelloEtAl2014}). The ISM abundance of [CO]/[H$_2$]
$\sim10^{-4}$ \protect\citep{AikawaNomura2006} is usually also assumed for discs.
However, it is important to note that there is a significant scatter around this value: \protect\cite{LacyEtAl1994} find a maximum value of the fractional abundance of $^{12}$CO of $9.1\times10^{-4}$, whereas \protect\cite{FrerkingEtAl1982} obtain a value of $\sim8.5\times10^{-5}$  in $\rho$~Oph and Taurus. 
Given the isotopic ratio  of [$^{16}$O]/[$^{18}$O] and its errors (557$\pm$30; \protect\citet{Wilson1999}), the resulting C$^{18}$O fractional abundance we employ is in a range between 
\begin{equation}
1.4 \times 10^{-7} < \frac{[\text{C}^{18}\text{O}]}{[\text{H}_2]} < 1.7 \times 10^{-6}\hspace{2pt}.
\label{eq:minmaxabund}
\end{equation}
The maximum effect of this uncertainty on the line emission is achieved in the optically thin case, where the line emission scales linearly with the abundance. We take this fully into account when presenting the results of our calculations of the C$^{18}$O line emission.
C$^{18}$O J=2--1 is excited by molecular collisions within the disc. 
The shape of its line is thus dependent on the temperature and density structure of the disc and on the C$^{18}$O mass available.

Using the \textsc{CASA} software package, we further process the data cubes to obtain synthetic \textsc{ALMA} observations that take into account filtering and instrumental and thermal noise effects. These can then be directly compared with or fitted to the observational C$^{18}$O data (e.g. molecular line profiles).
\pagebreak
\section{Results and discussion}
\label{subsec:models}
\subsection{Results of the SED modelling}
\label{subsec:SEDandSnowline}
   \subparagraph{\bf{Initial mass estimate}}
   We have explored in total over 100 models varying $M_\text{dust}$, $a_\text{min}$, $a_\text{max}$ and $g/d$ (alone, and in combination with $\alpha_\text{turb}$) and have found 15 models that fit the observed SED, the details of the models are given in Table \ref{tab:3models}. 
   
   In order to fit our model to the observed SED, we start from an initial estimate of the minimum dust mass, which determines the overall SED shape. If the resulting fluxes in the SED are too high compared to the observations, we decrease the dust mass. If the mm-wavelength fluxes are too high at shorter wavelengths, but not in the mm, we reduce the $g/d$ ratio. We then make further improvements on the fit by varying the minimum and maximum grain size, taking into account the effects of the individual disc parameters on the SED.
   For model series~B-E, we have based our initial dust mass estimate on the following considerations:
   for optically thin emission, the dust mass is given by
   \begin{equation}
   {M}_{\text{dust}}=\frac{{S}_\lambda {d}^2}{\kappa_\lambda {B}_\lambda({T})} \hspace{2pt},
   \label{eq:mdustmin}
   \end{equation}
   where $S_\lambda$ is the flux at a certain wavelength, $d$ the distance of the source in pc and ${\kappa_\lambda}$ the opacity at wavelength $\lambda$. $B_\lambda$ is the Planck function depending on the temperature, given by
   \begin{equation}
   {B}_\lambda= \frac{2h c^2}{\lambda^5} \left[\exp \left(\frac{h c}{\lambda k_\text{B}T} \right) -1   \right]^{-1} \hspace{2pt},
   \end{equation}
   where $h$ is the Planck constant and $k_\text{B}$ the Boltzmann constant. For a distance of $d\approx 120\,$pc \protect\citep{VanDenAnckerEtAl1998}, temperature $T\approx 20$\,K and at a wavelength of $\lambda \approx 0.85\,$mm, we find a flux density of $S \approx 2.1$\,Jy \protect\citep{GuidiEtAl2016}.
   We thus estimate a minimum dust mass of $M_\text{dust,min} \sim 8 \times 10^{-4}\,\text{M}_\odot$.
   Here, the opacity we assume is given in \protect\cite{Draine2006}, who shows that at $\lambda \approx1$\,mm, dust grains of size $a \approx 1$\,mm are most efficient emitters with $\kappa_\lambda \approx 4$\,cm$^2$ g$_\text{dust}^{-1}$ as they contain most of the mass (they are however not the most efficient emitters per unit mass).  Table~\ref{tab:3models} shows that this grain size is comparable to the maximum grain size $a_\text{max}$ of model series~C-E.
   These models therefore represent the case when the bulk of the mass is in $\sim 1$\,mm-sized grains and $M_\text{dust}$ is low.
   For a grain size of 0.4\,mm as in model series~B, $\kappa_\lambda$ from \protect\cite{Draine2006} is a little lower, leading to a slightly higher minimum dust mass than in model series~D to reproduce the same SED.
   Note, however, that we used the above calculation only to get an initial value for $M_\text{dust}$, employing standardized values for the opacity. The mm opacities $\kappa_\text{mm}$ we are using with \textsc{MCMax} are given in Table~\ref{tab:3models}, they depend on the dust grain size, thus they vary from model to model and differ slightly from the values given in \protect\cite{Draine2006}, which are for a material of specific chemical composition and properties assumed to be similar to ISM dust.
   However, the mm opacities only influence the exact location of the mm point in the SED.
     \begin{figure*}
     	\centering
     	\includegraphics[width=\textwidth]{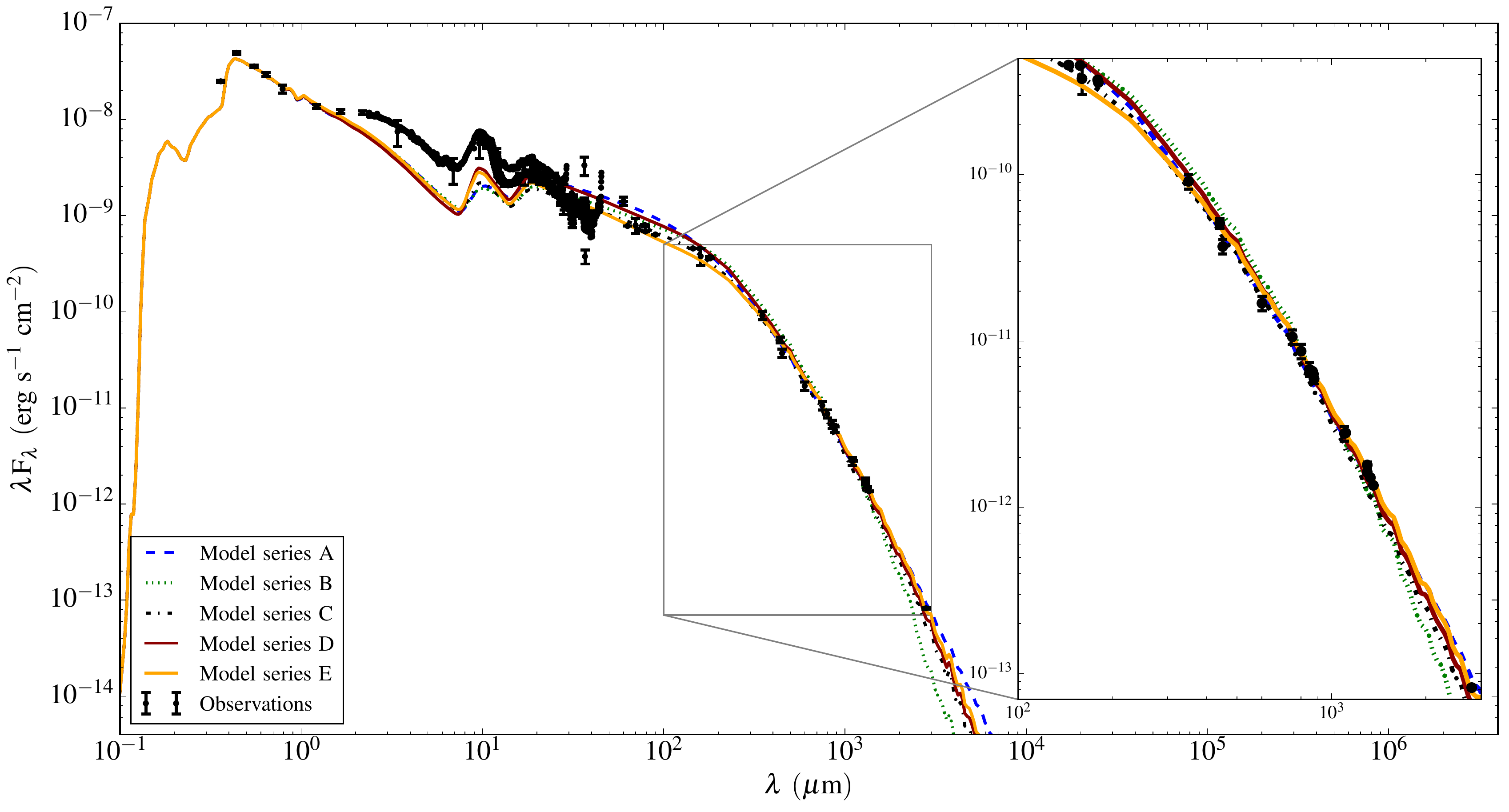}
     	\caption{
     		SEDs of our best-fitting models (parameters can be found in Table \ref{tab:3models}). Due to the degeneracy of $\alpha_\text{turb}$ and $g/d$, the SEDs within each model series are exactly the same (see text). The observations are plotted as black circles including the respective error bars. The observational data are taken from \protect\cite{BerrilliEtAL1992,ManningsSargent1997, BouwmanEtAl2000, IsellaEtAl2007, QiEtAl2011, TillingEtAl2012, MendigutiaEtAl2012, deGregorioMonsalvoEtAl2013} and the Spitzer c2d legacy survey.}
     	\label{fig:3SEDs}
     \end{figure*}
     \begin{center}
     	\begin{table*} 
     		\begin{tabular}{c c c c c c c c c c c}
     			\hline \bf{Model} & $M_{\text{dust}}$[M$_\odot$] & $M_{\text{gas}}$[M$_\odot$] & $g/d$ &  $a_{\text{min}}$[$\mu$m] & $a_{\text{max}}$[mm] & $T(90\,\text{au})$[K] & $p$ & $\alpha_\text{turb}$ & $\kappa_\text{mm}$[cm$^2$ g$_\text{dust}^{-1}$] & Distinctive features\\ 
     			\hline
     			\hline \bf{A} &  $3.0\times10^{-3}$ & $3.0\times10^{-1}$ & 100 & 0.8 & 35 & 25.0 & -1.0 & $10^{-4}$ & $\sim0.9$ & high $M_{\text{dust}}$ and $M_{\text{gas}}$\\ 
     			\hline \bf{B} &   $8.0\times10^{-4}$ &   $1.4\times10^{-1}$ & 180 & 0.8 & 0.4 & 22.5 & -1.0 & $10^{-4}$& $\sim4.0$& high $g/d$\\ 
     			\hline \bf{C} &  $8.0\times10^{-4}$ & $8.0\times10^{-2}$ & 100 & 0.5 & 1.1 & 22.0 & -1.0 &$10^{-4}$ & $\sim3.6$& intermediate-sized dust\\
     			\hline \bf{D} &   $\bf{9.5\times10^{-4}}$ & $\bf{1.0\times10^{-2}}$ & \textbf{10.5} & \textbf{0.02} & \textbf{1.4} & \textbf{23.5} & \textbf{-1.0}& $\bf{10^{-4}}$&$\bf{\sim3.3}$ & \bf{low $g/d$, pristine dust}\\  
     			\hline \bf{E} &  $\bf{1.3\times10^{-3}}$ & $\bf{1.2\times10^{-2}}$ & \textbf{9.2} & \textbf{0.05} & \textbf{1.4} & \textbf{20.0} & \textbf{-1.2} &$\bf{10^{-4}}$ & $\bf{\sim3.3}$& \bf{steeper $\Sigma(r)$-profile}\\ 
     			\hline 
     			\hline \bf{A/10} &  $\bf{3.0\times10^{-3}}$ & $\bf{3.0\times10^{-2}}$ & \textbf{10} & \textbf{0.8} & \textbf{35} & \textbf{25.0} & \textbf{-1.0} & $\bf{10^{-3}}$ & $\bf{\sim0.9}$ & \textbf{-}\\ 
     			\hline \bf{B/10} &   $\bf{8.0\times10^{-4}}$ &   $\bf{1.4\times10^{-2}}$ & \textbf{18} & \textbf{0.8} & \textbf{0.4} & \textbf{22.5} & \textbf{-1.0} & $\bf{10^{-3}}$& $\bf{\sim4.0}$& \textbf{-}\\ 
     			\hline \bf{C/10} &  $\bf{8.0\times10^{-4}}$ & $\bf{8.0\times10^{-3}}$ & \textbf{10} & \textbf{0.5} & \textbf{1.1} & \textbf{22.0} & \textbf{-1.0} &$\bf{10^{-3}}$ & $\bf{\sim3.6}$& \textbf{-}\\
     			\hline \bf{D/10} &   $9.5\times10^{-4}$ & $1.0\times10^{-3}$ & 1.05 & 0.02 & 1.4 & 23.5 & -1.0& $10^{-3}$&$\sim3.3$ &-\\ 
     			\hline \bf{E/10} &  $1.3\times10^{-3}$ & $1.2\times10^{-3}$ & 0.92 & 0.05 & 1.4 & 20.0 & -1.2 &$10^{-3}$ & $\sim3.3$& -\\ 
     			\hline 
     			\hline \bf{A/100} &  $3.0\times10^{-3}$ & $3.0\times10^{-3}$ & 1 & 0.8 & 35 &  25.0 & -1.0 & $10^{-2}$ & $\sim0.9$ & -\\ 
     			\hline \bf{B/100} &   $8.0\times10^{-4}$ &   $1.4\times10^{-3}$ & 1.8 & 0.8 & 0.4 & 22.5 & -1.0 & $10^{-2}$& $\sim4.0$& -\\ 
     			\hline \bf{C/100} &  $8.0\times10^{-4}$ & $8.0\times10^{-4}$ & 1 & 0.5 & 1.1 & 22.0 & -1.0 &$10^{-2}$ & $\sim3.6$& -\\
     			\hline \bf{D/100} &   $9.5\times10^{-4}$ & $1.0\times10^{-4}$ & 0.11 & 0.02 & 1.4 & 23.5 & -1.0& $10^{-2}$&$\sim3.3$ & -\\ 
     			\hline \bf{E/100} &  $1.3\times10^{-3}$ & $1.2\times10^{-4}$ & 0.09 & 0.05 & 1.4 & 20.0 & -1.2 &$10^{-2}$ & $\sim3.3$& -\\ 	     			
     			\hline 
     		\end{tabular}  
     		\caption {Parameters of our 15 models that fit the observed SED: $M_{\text{dust}}$, $M_{\text{gas}}$, $g/d$, $a_{\text{min}}$, $a_{\text{max}}$ and $T_{\text{mid-plane}}$ at the location of the CO snowline radius $R_\text{sl}=90\,$au. We also give the power-law exponent $p$ of the surface density profile $\Sigma \propto r^{p}$. We list as well the turbulent mixing strength $\alpha_\text{turb}$ and the mm opacity $\kappa_\text{mm}$ of the dust grains that \textsc{MCMax} is using. The respective distinctive characteristics of the models are given in the last column. Models~(A-E)/10 and (A-E)/100 have the same dust properties as models~A-E, but their $g/d$ (and thus their $M_\text{gas}$) are divided by factors of 10 (100) and their $\alpha_\text{turb}$ multiplied by 10 (100) in comparison with models~A-E. The models given in boldface are the models that also match the observed C$^{18}$O line profiles as we will discuss in the next section. }
     		\label{tab:3models} 
     	\end{table*}
     \end{center} 
\subparagraph{\bf{Our models}}
The SEDs of all our models are given in Figure~\ref{fig:3SEDs}. As discussed above, all models of a given model series have the same SED due to the degeneracy of $\alpha_\text{turb}$ and $g/d$. We include a zoom-in of the far-infrared (FIR) and mm region of the SED as this is the wavelength regime that is most crucial for our analysis. All our models match the observed SED in this wavelength range very well, therefore the different models are hardly distinguishable there.
We do not try to fit the observed SED at wavelengths $\lambda \gtrsim 2$\,mm because emission in this regime can be dominated by free-free-emission \protect\citep{WrightEtAl2015,GuidiEtAl2016}, which is not included in our calculations. 
Note that the only models which can reproduce the $\lambda >2$\,mm observations by \textit{thermal} emission - and no free-free emission at all - are the models of series~A, D and E, where series A needs large grains (35\,mm) and all three of them a high dust mass.
Furthermore, the models do not match the near-infrared (NIR) excess at $\lambda<10\,\mu$m very well, but this is sensitive to the exact dust grain composition and geometry of the very inner disc \protect\citep[inner few au;][]{Meijer2008}.
Including a puffed-up inner disc rim, which could potentially cast a shadow on to the disc surface,  might be expected to provide a better fit to the NIR SED. However \protect\cite{AckeEtAl2009} find that this would only influence the NIR regime of the SED (not the FIR or mm). Therefore a puffed-up inner rim would not improve the fit over a substantial wavelength range. The NIR fit does not alter our results for the dust mass, which is calculated using the longer wavelength component of the SED. Furthermore, adjusting the scale-height in the inner disc would violate the self-consistency of our models. For simplicity, we therefore do not include a model for a puffed up inner rim at this stage.

 \subparagraph{\bf{Description of model series (A, B, C, D, E)}}
 We will first discuss the general characteristics of each of these models.
 We find that model~A has a relatively high $M_\text{dust}$ and $M_\text{gas}$ (of the order of 10$\%$ of the stellar mass $M_*=2.3$M$_\odot$). This will in general produce relatively high fluxes, so in order to compensate for the high masses, its grains have to be quite evolved. This will ensure that not too many small grains are situated in the disc atmosphere, where they could intercept stellar irradiation and thus boost the fluxes of the SED.
 Models~B and C have the same $M_\text{dust}$, but due to their different $g/d$ ratios, their $M_\text{gas}$ values differ by about a factor of 2. In order for them to give the same SEDs, model~B with the higher $g/d$ has to have different grain sizes from model~C: its minimum grain size is a little larger than for model~C, while its maximum grain size is about a factor of 3 smaller. The distinctive feature of model~B is its high $g/d$ ratio which we compensate for by making its minimum grain size bigger than in model~C. Therefore - as described for model~A - the fluxes in the SED are reduced.
 Models~D and E have a much lower gas mass than the other models we found, but they all have the same SED. This is caused by these two models being the only ones with very small grains. These are coupled to the gas, dragged to higher layers and thus intercept more stellar flux. In general, models~A-C have relatively large grain sizes (which probably corresponds to a more evolved state), otherwise the SEDs would produce too high values in the FIR.
 Model~E is similar to model~D, but employs a different radial dependence of the surface density: for models series~A-D we assumed $\Sigma \propto r^{-1}$, for model series~E we take a slightly steeper profile of $r^{-1.2}$, although still well within the range of observationally measured values for protoplanetary discs \protect\citep{AndrewsEtAl2010}. Model series~E yields a lower mid-plane temperature at the location of the CO snowline radius than the other models, as we will discuss shortly.
 
We have calculated the optical depth of the continuum emission for our models (using the surface density and mm opacity), which yields that the mm continuum emission for all our models is optically thin, except for model series~E, where it becomes optically thick within the inner $\sim10$\,au. This enables us to obtain a reliable estimate of the dust mass in the disc. We would like to highlight that some of our models reach the maximum mm opacity (as obtained by \protect\citealt{Draine2006}) and are indicative of the minimum $M_\text{dust}\sim8\times10^{-4}$M$_\odot$ as derived earlier in this section. Much lower $\kappa_\text{mm}$ are of course possible if a big fraction of the mass is hidden in pebbles and larger bodies, which do not contribute to the mm flux. In such cases, $M_\text{dust}$ in our models is just indicative of the mass of the mm dust and thus a much higher total mass of solids can be achieved. However, such models would not differ in the SED.
 
 \subparagraph{\bf{Variations of model series (A-E/10, A-E/100)}}
   Models~A-E have a turbulent mixing strength of $\alpha_\text{turb}=10^{-4}$, but as described above we run additional calculations for these models, exploring larger values of $\alpha_\text{turb}$ ($10^{-3}$ and $10^{-2}$) while decreasing the gas masses in these models by factors of 10 and 100. 
  We leave all the remaining parameters of models A-E unchanged, as listed in Table~\ref{tab:3models}.
  Indeed, we find that we can match the observational constraints given by the SED and mid-plane temperature requirements by all our models A-E, by changing the $M_\text{gas}$ and adjusting the $\alpha_\text{turb}$ accordingly, to keep $M_\text{gas}\times\alpha_\text{turb}$ (and therefore the dust diffusion solution, resulting temperature structures and the SEDs within each model series) constant. 
  In general, all models A-E/10 and A-E/100 yield low to very low $g/d$ ratios by construction. In order to compensate for the lower gas masses, higher levels of turbulent mixing are needed to transport the dust grains to higher altitudes in the disc where they can absorb the stellar light and give the same SED.
 Given that the dust grain properties within a certain series are exactly the same, our above description of the distinctive features of models~A-E holds true for (A-E)/10 and (A-E)/100, respectively.

 In general, a surprising result of our modelling is that we can match both the SED and CO snowline radius, by making very different assumptions on the basic parameters, such as dust grain size and gas mass.
 Higher emission can for example be caused by a higher dust or gas mass, but also by smaller dust grains present in the disc. Some of these parameter degeneracies are also discussed in \protect\cite{Meijer2008, WoitkeEtAl2015} and Pani{\'c} et al. (subm.).
It is therefore important to note that SED modelling alone does not provide unambiguous physical models of the disc structure, but is highly degenerate.

   \subsection{{Disc regions determining the SED}}
\begin{figure}
      	\centering
      	\includegraphics[width=\columnwidth]{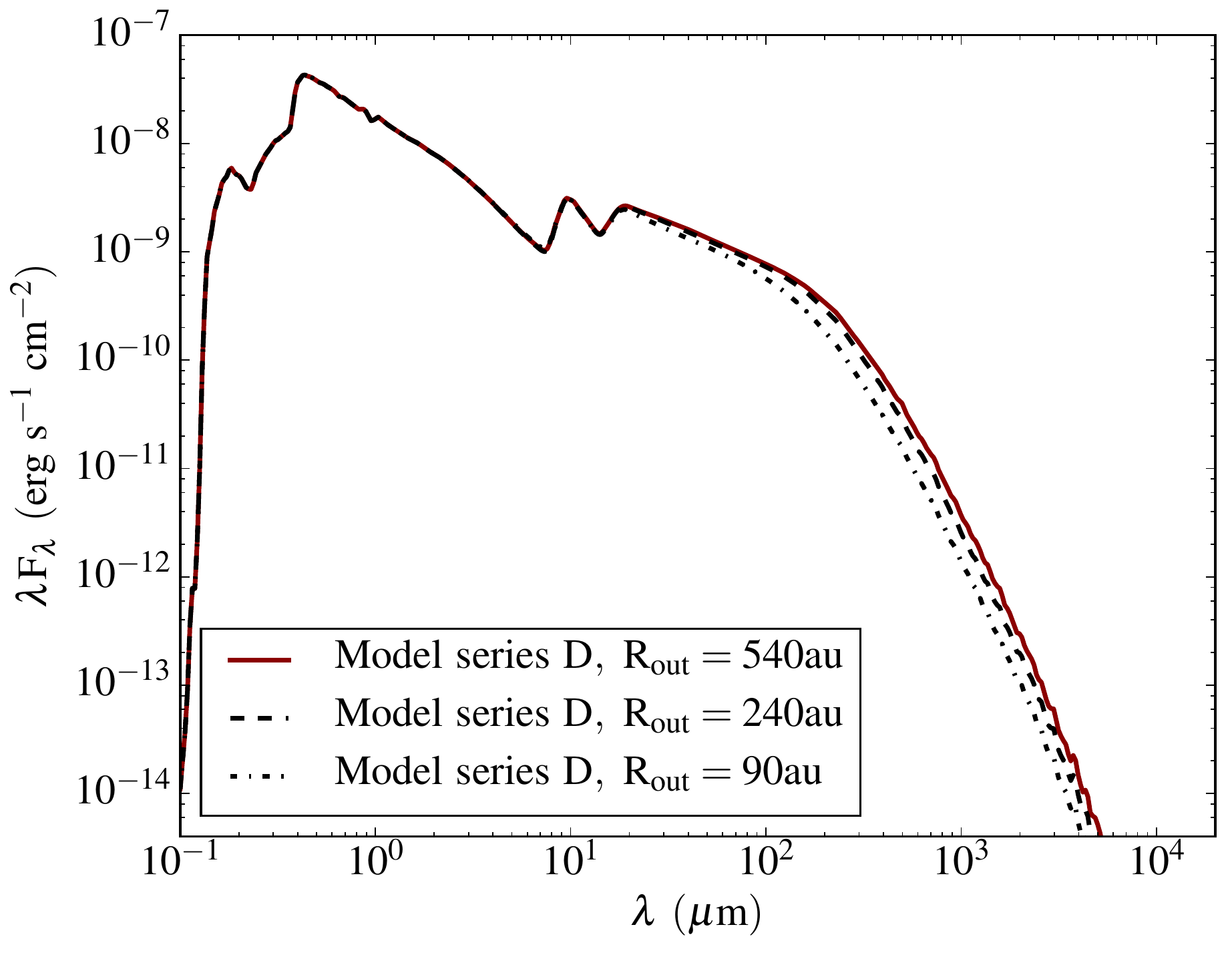}
      	\caption{SED of the whole disc $R<540\,$au (red solid line), from within $R<R_{850\,\mu\text{m dust}}=240\,$au (black dashed line) and from $R<R_\text{sl}=90\,$au (black dash-dotted line). }
      	\label{fig:SEDinner}	
      \end{figure}

   Fitting observed SEDs in general is most suited for the inner disc regions, since dust grains in the outer disc regions do not intercept sufficient stellar light to contribute substantially to the SED.
   We have checked that the SEDs as obtained from the whole discs ($R=540$\,au) of our 15 models (as shown in Figure~\ref{fig:3SEDs}), are only marginally changed when taking the emission from within $240\,$au. This is plotted in Figure~\ref{fig:SEDinner} for model series~D, we find the same behaviour for the other models.
   Thus the fact that our model discs are described by a single power-law surface density distribution out to 540\,au (whereas
   the observed dust distribution extends only to $\sim240$\,au)
   will not have a significant effect on our SED fits. We will
   later focus on the gas budget of the disc within the CO snowline
   ($R_\text{sl}=90$\,au) and note that, in our models, this region contributes
   around $40-50\%$ of the flux at sub-mm wavelengths. We also overplot the SED for the emission from within the snowline radius in Figure~\ref{fig:SEDinner}.

   \subsection{{Models matching the CO snowline location}}
 \begin{figure}
 	\centering
 	\includegraphics[width=\columnwidth]{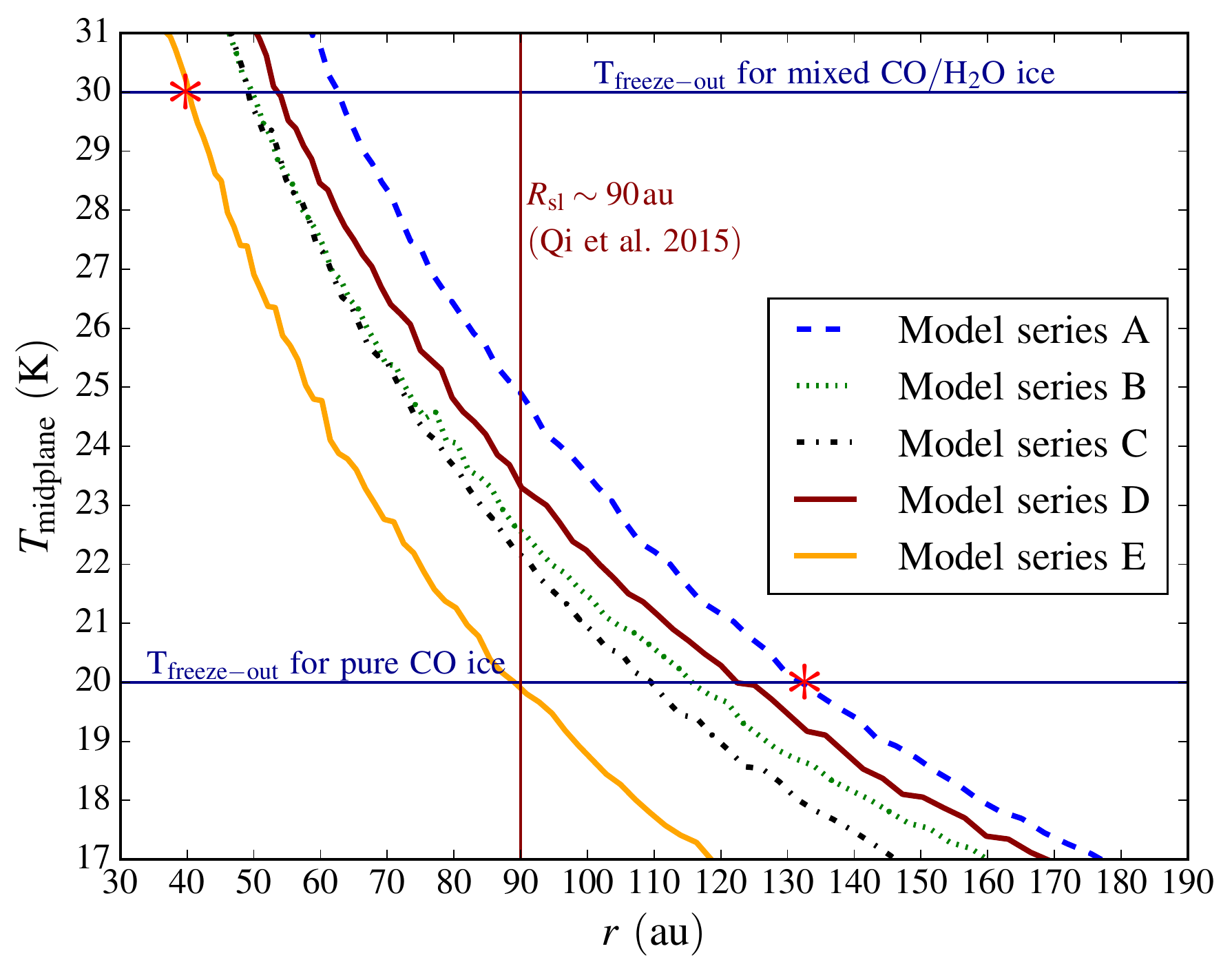}
 	\caption{Mid-plane temperature as a function of radius for our 15 disc models that match the observed SED. Models from the same series have exactly the same mid-plane temperature structure (see Section~\ref{subsec:2dstructuremcmax}). In dark red (vertical) we plot the observed snowline radius $\approx 90$\,au \protect\citep{QiEtAl2015}.
 	The upper and lower limits of the freeze-out temperature ($\sim 20-30\,$K) as found by e.g. \protect\cite{CollingsEtAl2004} are plotted as horizontal lines.
 	The red asterisks indicate where the snowline location could be between $\sim$40 and 135\,au due to a plausible range of freeze-out temperatures of 20-30\,K if the snowline location was not known unambiguously from observations.}
 	\label{fig:Tmidplane}	
 \end{figure}
 
 As an additional constraint we have to make sure that our \textsc{MCMax} models are consistent with the observed CO snowline location. Therefore we analyse the mid-plane temperature profile $T_{\text{mid-plane}}(r)$ for all our models that match the observed SED, which we plot in Figure~\ref{fig:Tmidplane}. As mentioned above, models of the same series have the same temperature structure.
 We find that all of them have mid-plane temperatures between $\sim 20$ and $25\,$K at the observed snowline radius of $R_\text{sl} \approx \nolinebreak 90$\,au \protect\citep{QiEtAl2015}. These are well within the range of values generally assumed and observed for the freeze-out temperature of CO:
The freeze-out temperature can vary between $\sim20$ and $\sim$30\,K depending on whether CO is binding to pure CO ice or a mixture with water ice \protect\citep{CollingsEtAl2004}, which is, in turn, also dependent on the chemical history of the ice \protect\citep{GarrodPauly2011}.
In general, the CO freeze-out temperature is not known unambiguously and might vary from system to system \protect\citep{HersantEtAl2009,QiEtAl2015}:
 \protect\cite{QiEtAl2013} found a freeze-out temperature of CO of 17\,K from their modelling of TW Hya, whereas  \protect\cite{JorgensenEtAl2015} obtain temperatures of about 30\,K in their study of embedded protostars. 
  \protect\cite{QiEtAl2011} assume a freeze-out temperature for CO of $T\approx 19\,$K (pure CO ice) for HD~163296. However, in  \protect\cite{QiEtAl2015} they perform a new analysis with higher-resolution observational data and use a temperature in the mid-plane at $R_\text{sl}$ of $T\approx 25\,$K (mixed CO/H$_2$O ice). 
Thus all our models have temperatures in the disc mid-plane at the location of the snowline radius that are well within the plausible range. Our exploration of self-consistent models confirms that all the freeze-out temperatures assumed in these previous literature references fall within the plausible range of temperatures for HD~163296. If the freeze-out temperature of CO was known unambiguously, this would, in combination with an observationally determined CO snowline location, be a powerful model discriminant and we might be able to exclude models based on this constraint.
Since, however, it is unclear what exactly the relevant freeze-out temperature is, we find that all the models can match the observed snowline location of 90\,au.
This weak model discrimination also means that it is impossible to {\it predict} the CO snowline radius from SED model fitting alone or even from fitting the molecular line emission together with the SED \protect{\citep{QiEtAl2011}}: given the uncertainty in the sublimation temperature of CO,
our viable SED fits imply predicted radii in the range $\sim$40-135\,au as
denoted by the red asterisks in Figure~\ref{fig:Tmidplane}.
In general, we find that the location of the CO snowline radius does not further discriminate between models in comparison to the criterion given by the SED; however it sets the radial location inwards of which no freeze-out is taking place in our models, and which is therefore important for the interpretation of the C$^{18}$O emission. It is important to note that an SED fit does not determine the CO snowline location and that, vice versa, a CO snowline observation does not discriminate amongst possible SED models.

 We conclude that all our 15 models match the SED and CO snowline location within the uncertainties in the freeze-out temperature.
 The SED modelling is especially powerful for the inner $\sim240\,$au and describes the disc structure inside the CO snowline location well.
   
 \subsection{C$^{{18}}$O J=2-1 emission}
 \label{subsec:c180_emission}
 \subsubsection{C$^{{18}}$O line profiles}
\begin{figure}
        	\centering
        	\includegraphics[width=\columnwidth]{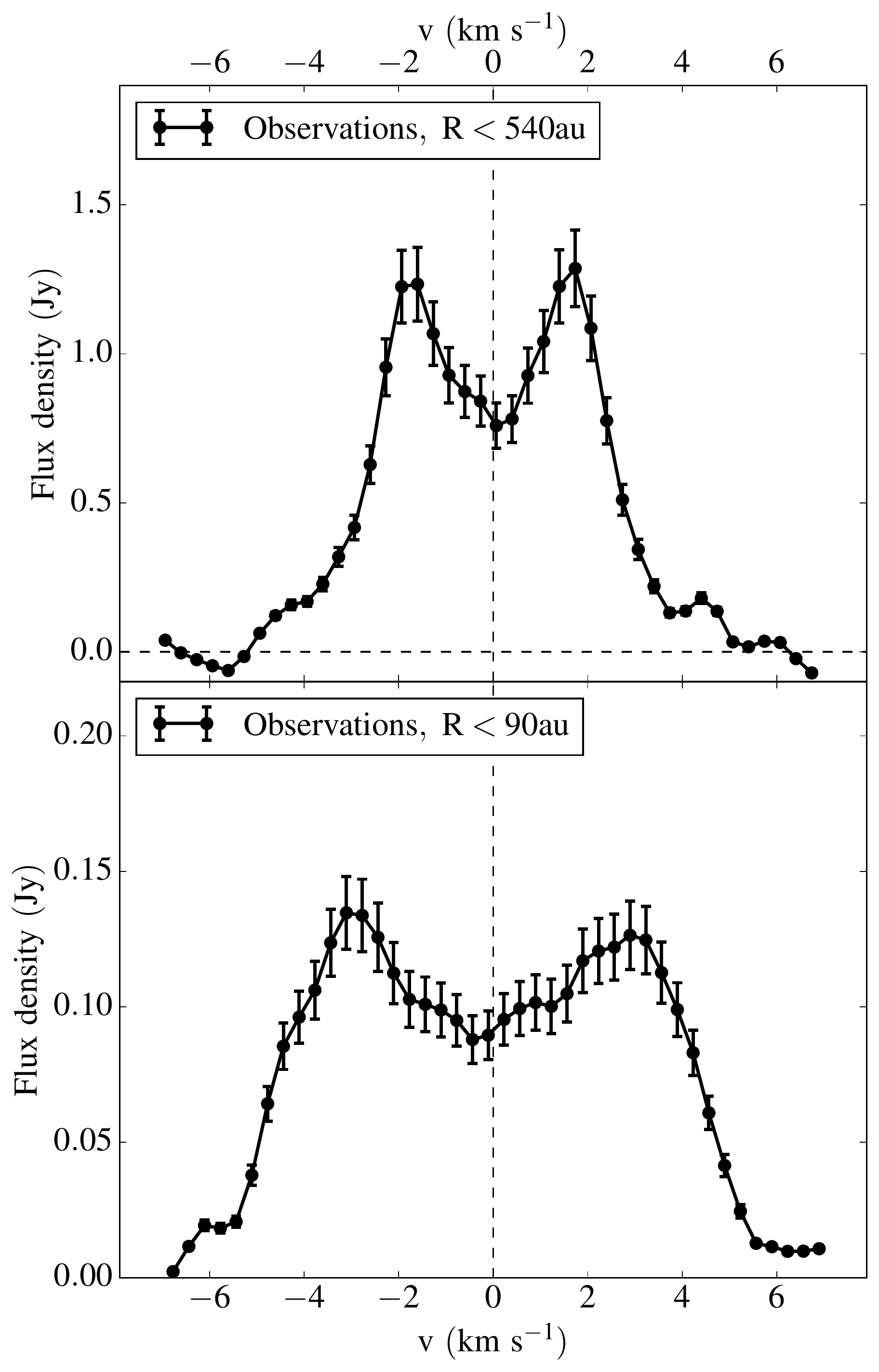}
        	\caption{C$^{18}$O J=2-1 line profile (ALMA observations) using the emission from the whole disc (upper panel) and from within the 90\,au snowline radius. The error bars represent a $10\%$ flux calibration uncertainty \protect\citep{GuidiEtAl2016}. We centred the spectra on 0\,km\,s$^{-1}$. The systemic velocity is $v_\text{sys}$=5.8\,km\,s$^{-1}$. }
        	\label{fig:line_profile}
        \end{figure}

 As discussed in Section~\ref{subsubsec:torusAlma}, we use the density and temperature structures of our \textsc{MCMax} models to calculate the C$^{18}$O line emission with \textsc{torus}. 
 The aim of our work is to interpret the observed C$^{18}$O J=2-1 line profile,
 especially from the inner disc regions, in the context of a physically consistent model that matches other relevant observations ($R_\text{sl}$, SED).
We show the C$^{18}$O J=2-1 line profile as observed by ALMA in Figure~\ref{fig:line_profile}, both taking the emission from the whole disc and from within the 90\,au snowline radius. These are very different, especially in the peaks of the spectrum, as these are dominated by the emission from the outer disc regions. Our goal is to model the disc regions with $R<90\,$au (the snowline location), as these are independent of the details of freeze-out in the outer disc.

 \subparagraph{\bf{Uncertainty in abundance of C$^{18}$O}}
 As mentioned previously, the fractional abundance of C$^{18}$O has a large uncertainty and is observed (in star-forming clouds) to be in a range between $1.4\times10^{-7}$ and $1.7\times10^{-6}$ (see Section~\ref{subsubsec:torusAlma}), thus spanning an order of magnitude. From matching the observed C$^{18}$O J=2-1 line profile with our models, we can unambiguously calculate the mass of C$^{18}$O in the disc. However, when then converting this mass to a mass of H$_2$ we will have to take into account this range of C$^{18}$O abundance.
 
 Matching the observed C$^{18}$O line profile, we find that - taking into account the range of plausible abundances - only five of our models can fulfil this criterion, namely (A-C)/10, D and E. We will thus focus on these models in the further discussion.

Another aspect to take into account is that the abundance of C$^{18}$O in comparison to H$_2$ can be altered due to freeze-out in the disc mid-plane, as already discussed in the previous section. We have implemented this effect in \textsc{torus} by setting the abundance of C$^{18}$O to a negligible value when the disc temperature drops below the freeze-out temperature. For the individual models, we use the mid-plane temperature of the respective model at the observed snowline radius as given in Figure~\ref{fig:Tmidplane}.
The fraction of the C$^{18}$O mass removed by freeze-out in the respective models is given in Table~\ref{tab:percentagesmodels}. We find that this effect is stronger for models~(A-C)/10 than for models~D and E, because the former have slightly higher gas masses and bigger grains, thus a higher fraction of the mass will be concentrated in the cooler disc mid-plane regions and thus subject to freeze-out. However, the exact impact of freeze-out on the line profile will depend on the details of the vertical temperature profile and thus on the location of the CO ice surface.
 
\begin{figure}
        	\centering
        	\includegraphics[width=\columnwidth]{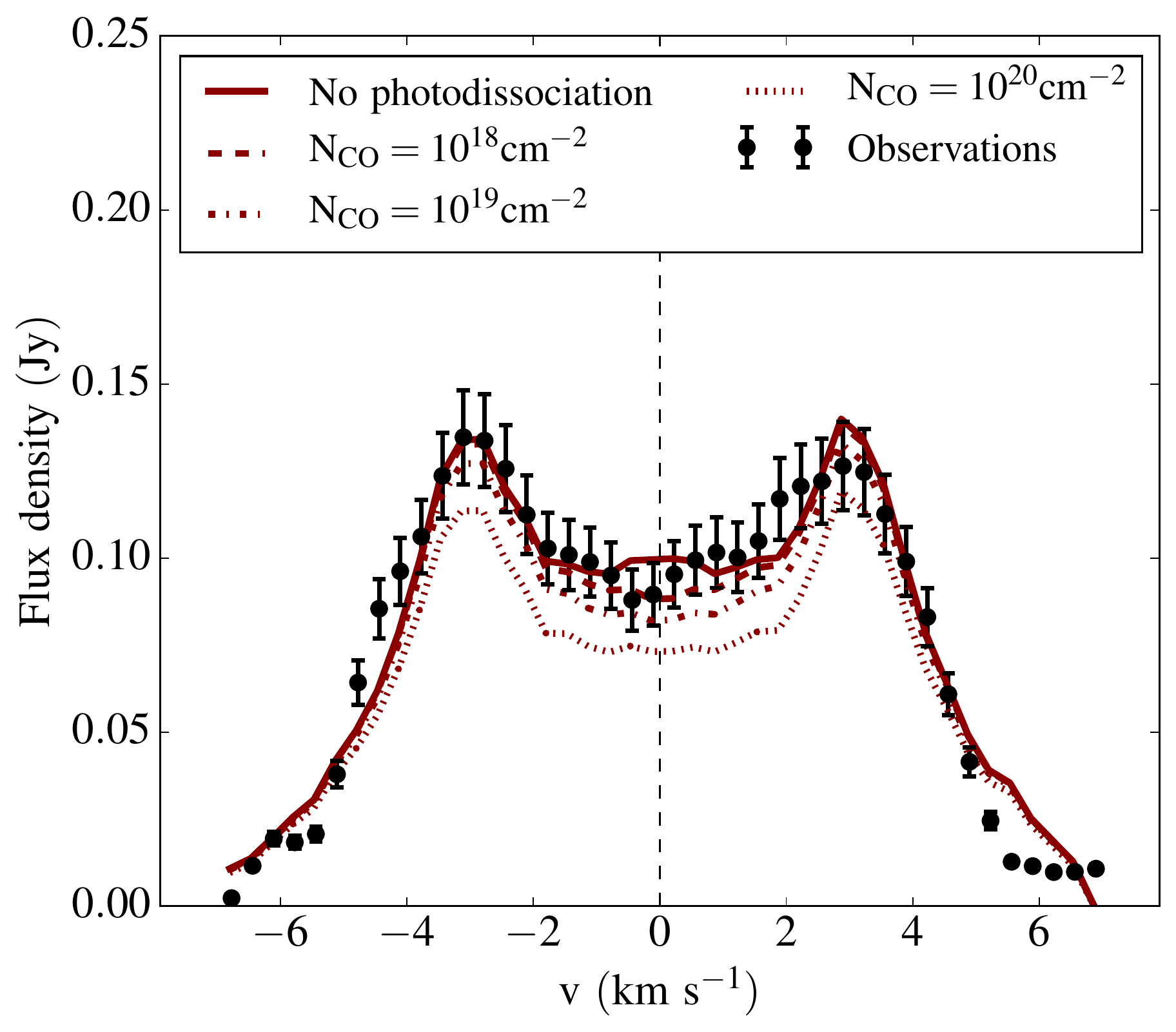}
        	\caption{C$^{18}$O flux density for model~D (from regions with $R<R_\text{sl}=90\,$au) employing different column density thresholds for photodissociation: no photodissociation (solid), $N_\text{CO}$=$10^{18}$\,cm$^{-2}$ (dashed), $10^{19}$\,cm$^{-2}$ (dash-dotted) and $10^{20}$\,cm$^{-2}$ (dotted). Even in the most extreme case, only $\sim 3\%$ of the CO mass is affected. For reference, we also plot the observed line profile for this disc region, given by the black points with error bars ($10\%$ flux calibration uncertainty). We find a very similar behaviour for the other models and thus do not show the respective plot here.}
        	\label{fig:photodiss}
        \end{figure}

The second most relevant source of CO-removal from the gas-phase is photodissociation \mbox{\protect\citep{VisserEtAl2009,MiotelloEtAl2014}}. As described in Section~\ref{subsubsec:torusAlma}, we have taken this into account in our modelling. 
We do this by setting the C$^{18}$O abundance to a negligible value for a threshold column density of gas calculated from the star in different azimuthal directions covering the entire disc height. These threshold column densities vary from $N_\text{CO}$=$10^{18}$\,cm$^{-2}$, $10^{19}$\,cm$^{-2}$ and $10^{20}$\,cm$^{-2}$ (corresponding to $10^{22}$\,H$_2$\,cm$^{-2}$, $10^{23}$\,H$_2$\,cm$^{-2}$ and $10^{24}$\,H$_2$\,cm$^{-2}$, respectively, assuming $f_\text{CO}\sim 10^{-4}$).
      \begin{center}
      	\begin{table}
      		\begin{tabular}{c c c c c c   }
      			\hline \bf{Model series}& \bf{A} & \bf{B} &  \bf{C} & \bf{D} & \bf{E} \\ 
      			\hline
      			$T_\text{freeze-out}$(90\,au) [K] & 25.0 & 22.5 & 22.0 & 23.5 & 20.0 \\ 
      			\hline& $\sim$48$\%$  & $\sim$45$\%$ & $\sim$40$\%$ & $\sim$27$\%$ & $\sim$23$\%$  \\ 
      			 \hline 
      		\end{tabular}  
      		\caption {Freeze-out temperatures (mid-plane temperatures at 90\,au) and fractions of the CO mass removed in the various model series due to freeze-out.}
      		\label{tab:percentagesmodels} 
      	\end{table}
      \end{center}
We find that overall only a small fraction of C$^{18}$O is photodissociated within the 90\,au snowline location in our models. For the first threshold, the fraction of the CO gas mass photodissociated is $\sim0.1\%$ in all our models, for $N_\text{CO}=10^{19}$\,cm$^{-2}$ it is $\sim1\%$ and even for the extreme case, only $\sim3\%$ is photodissociated within the snowline radius.
    We calculate the C$^{18}$O line emission for $R<90\,$au after removing CO from the photodissociated layer, in the three explored cases ($N_\text{CO}$=$10^{18}$\,cm$^{-2}$, $10^{19}$\,cm$^{-2}$ and $10^{20}$\,cm$^{-2}$).
    This is presented for the example of model~D in Figure~\ref{fig:photodiss}.
Given that in our analysis we mostly focus on the regions within the 90\,au snowline radius that are not subject to freeze-out and not strongly affected by photodissociation, our models are not dependent on the exact details of these processes.

\begin{figure*}
	           	\centering
	           	\includegraphics[width=0.77\textwidth]{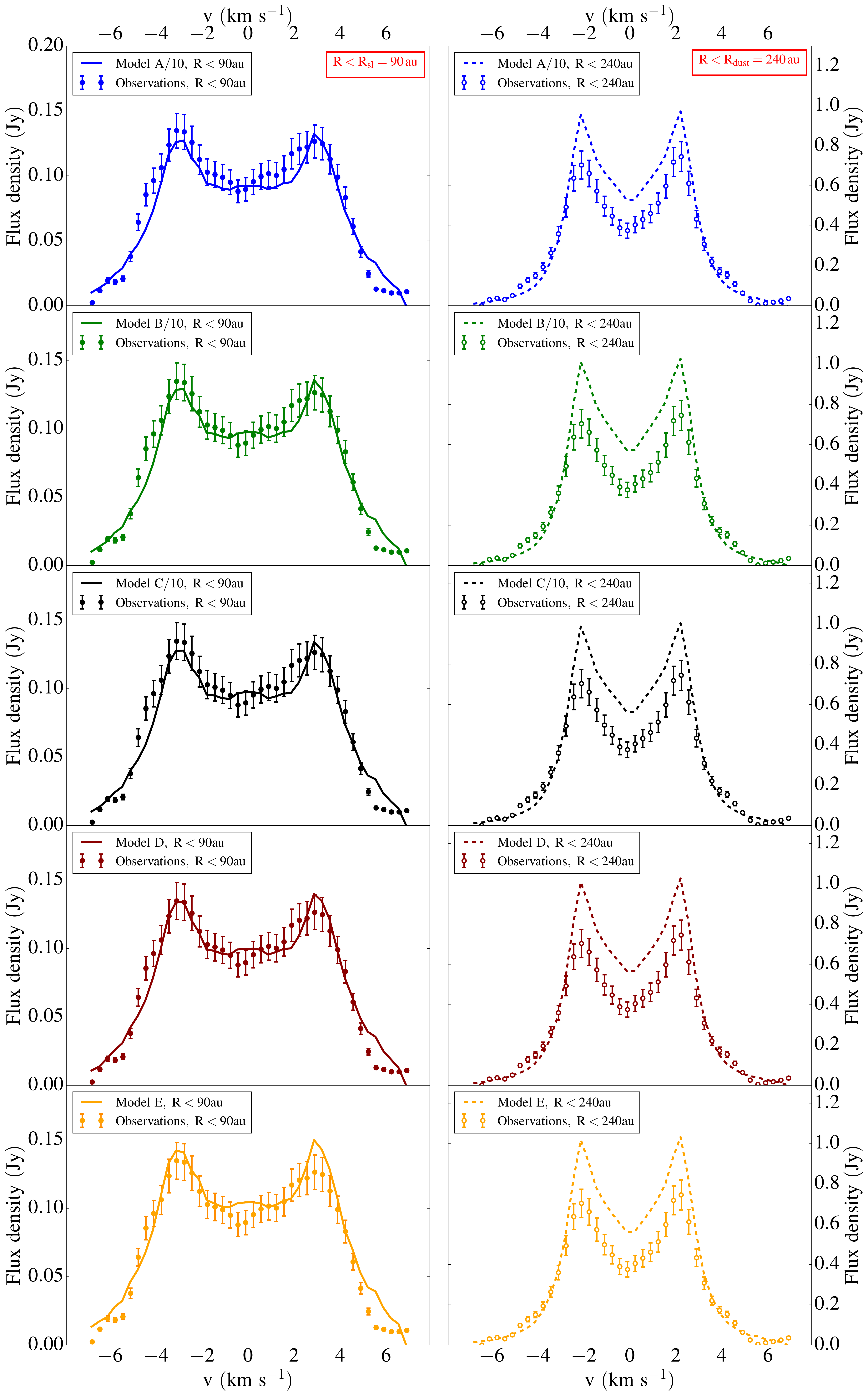}
	           	\caption{C$^{18}$O J=2-1 line profiles for our five models: We show the emission from within the CO snowline radius (90\,au) as well as from a disc region with R<240\,au (this corresponds to the outer disc radius as obtained from mm continuum observations). The line profiles from our models are given by the lines (solid for R=90\,au, dashed for R=240\,au), the spectra of the observations by the respective dots. The error bars reflect the $10\%$ flux calibration uncertainty of the observations. All models of the same series that have abundances in the range of C$^{18}$O abundances we consider will have the same flux densities because the C$^{18}$O masses and temperature structures are the same for each of these model series. The profiles were centred around 0\,km\,s$^{-1}$ (the systemic velocity is $v_\text{sys}$=5.8\,km\,s$^{-1}$).}
	           	\label{fig:spectra_10}
	           \end{figure*}

\subparagraph{\bf{The inner disc regions ($\bf{R<90}\,$au)}}
For 5 out of our 15 initial models, we can match the observed C$^{18}$O J=2-1 line profile within the range of plausible C$^{18}$O abundances between $1.4\times10^{-7}$ and $1.7\times10^{-6}$ (Equation~\ref{eq:minmaxabund}). We show the C$^{18}$O spectra arising from the regions inside the 90\,au snowline radius in these models in the left-hand panels panels of Figure~\ref{fig:spectra_10}. We will call these five models (A-C)/10, D and E "fiducial" models in the further discussion.
We obtain these from the synthetic and ALMA data cubes using the \textsc{casa} software in the following way: we calculate the emission from en elliptical region, centred on the centre of the disc using a PA=132$^{\circ}$ and a ratio of minor to major axis $\frac{b}{a}=\cos{i}$, where $i=48^{\circ}$. These values of PA and inclination are the ones we obtained in the analysis of the observations (see Section~\ref{sec:observations}).
All of them match the observations well within the error bars (given by the $\sim 10\%$ flux calibration uncertainty of the ALMA observations, \protect\cite{GuidiEtAl2016}). 
The fractional abundances of C$^{18}$O of these models can be found in Table~\ref{tab:90au}. Given these abundances and the gas masses in the respective models, we can calculate the mass of C$^{18}$O within the snowline radius, as the C$^{18}$O J=2-1 transition is mainly optically thin throughout the whole disc. 
We have calculated the optical depth of the C$^{18}$O J=2-1 transition and found that it is indeed optically thin throughout the whole line profile and at all radii for all our models that match the observations. Although the models are optically thin, this would not have been a necessary precondition for our modelling process as the radiative transfer calculation self-consistently accounts of optical depth effects. 
However, the low optical depths emphasizes how essential C$^{18}$O is as a tracer for the disc mid-plane.
That implies that we can unambiguously calculate the ${M}_{\text{C}^{18}\text{O}}$ within 90\,au which should be approximately the same for all our models. The values we obtain are listed in Table~\ref{tab:90au} and are in a range of 
\begin{equation}
{M}_{\text{C}^{18}\text{O}}(R<90\,\text{au}) \approx 2-3 \times 10^{-8} \text{M}_\odot \hspace{2pt}.
\label{eq:mc18o_90au}
\end{equation}
The values that \protect\cite{QiEtAl2011} obtain for these inner disc regions are comparable to ours.
In the right-hand panels of Figure~\ref{fig:spectra_10}, we plot the emission from a bigger disc region, namely from within the outer dust radius $R_\text{dust}\approx240\,$au. Our models still closely match the wings of the spectrum and thus the emission from the inner disc regions. However, one can see that our models slightly over-predict the emission from the outer disc regions (i.e. in the peaks of the spectrum) there. The height of the peaks depends  crucially on the exact vertical temperature structure of the models as freeze-out will reduce the C$^{18}$O emission, especially in the outer disc regions. However, we do not attempt to match these disc regions, but focus on the innermost 90\,au. 
\begin{center}
	\begin{table*}
		\begin{tabular}{c c c c c c   }
			\hline & \bf{A/10} &  \bf{B/10} &  \bf{C/10} & \bf{D} & \bf{E} \\ 
			\hline
			\hline 
			${f}_{\text{C}^{18}\text{O}}$& $2.7\times10^{-7}$ &  $6.4\times10^{-7}$ &  $1.0\times10^{-6}$ & $6.6\times10^{-7}$ & $6.8\times10^{-7}$\\
			\hline
			${M}_{\text{C}^{18}\text{O}}$($R<90$\,au) [M$_\odot$] & $1.9\times10^{-8}$ & $2.1\times10^{-8}$ &  $1.9\times10^{-8}$ & $1.6\times10^{-8}$ & $2.7\times10^{-8}$ \\ 
			\hline ${M}_{\text{gas}}$($R<90$\,au) [M$_\odot$] & $5.0\times10^{-3}$ &  $2.3\times10^{-3}$ & $1.3\times10^{-3}$ & $1.7\times10^{-3}$ & $2.9\times10^{-3}$ \\
			\hline ${M}_{\text{dust}}$($R<90$\,au) [M$_\odot$] & $5.0\times10^{-4}$ & $1.3\times10^{-4}$ & $1.3\times10^{-4}$ & $1.6\times10^{-4}$& $3.1\times10^{-4}$ \\
			\hline $g/d$ & 10 & 18 &  10 & 10.5 & 9.2\\
			\hline $\alpha_\text{visc}$($R<90$\,au) & 0.2 &  0.4 &  0.7 & 0.5 & 0.3\\
			\hline
		\end{tabular}  
		\caption {Properties of our five best-fitting (fiducial) models: fractional abundance of C$^{18}$O as obtained by matching the observed line profile within the 90\,au snowline radius, the mass of C$^{18}$O within this radius, the H$_2$ mass within the snowline radius, the average $g/d$ within 90\,au and $\alpha_\text{visc}$ for the respective cases. }
		\label{tab:90au} 
	\end{table*}
\end{center} 

\subparagraph{\bf{Models with minimum and maximum $\mathbf{g/d}$}}
So far, we have only explored the five models from our initial Table~\ref{tab:3models} that also match the C$^{18}$O line profile. However, it is interesting to look into the extreme cases, i.e. models with minimum and maximum plausible C$^{18}$O abundance (and thus maximum and minimum $g/d$ and $M_\text{gas}$), while still matching the observed C$^{18}$O and thus the C$^{18}$O masses within 90\,au we just calculated. We give the properties of the extreme models in Table~\ref{tab:extreme}.
It is important to note that models~D$_\text{max}$ and E$_\text{max}$ can be excluded as their $\alpha_\text{turb}$ is lower than the minimum of $10^{-4}$ we assume. Models D and E both have this minimum value; therefore for model series D and E, the highest possible values of $g/d$ and therefore $M_\text{gas}$ within 90\,au are the ones given in Table~\ref{tab:90au}. The highest possible values of $g/d$ for models that match the observed C$^{18}$O line profiles are 82 and 71 (for models~ B$_\text{max}$ and C$_\text{max}$, respectively); the lowest value is 2 (model~A$_\text{min}$). We take these three cases into account for the further discussion as they are the extreme ends of the $g/d$ range we obtain.
\begin{center}
	\begin{table*}
		\begin{tabular}{c c c c c c c c } 
			\hline & ${f}_{\text{C}^{18}\text{O}}$&  ${M}_{\text{C}^{18}\text{O}}$($R<90\,$au) [M$_\odot$] & ${M}_{\text{gas}}$($R<90\,$au) [M$_\odot$] & ${M}_{\text{dust}}$($R<90\,$au) [M$_\odot$]& $g/d$ &  $\alpha_\text{turb}$ &  $\alpha_\text{visc}$($R<90\,$au) \\
			\hline \hline
			\bf{A$_\text{min}$}& $1.7\times10^{-6}$ & $1.9\times10^{-8}$ & $7.9\times10^{-4}$ & $5.0\times10^{-4}$ & 2 & $6.3\times10^{-3}$ & 1.2 \\
			\hline
			\bf{A$_\text{max}$} & $1.4\times10^{-7}$ & $1.9\times10^{-8}$ & $9.6\times10^{-3}$ & $5.0\times10^{-4}$ & 19 & $5.3\times10^{-4}$ &  0.1\\ 
			\hline \hline
			\bf{B$_\text{min}$}& $1.7\times10^{-6}$ & $2.1\times10^{-8}$ & $8.7\times10^{-4}$ & $1.3\times10^{-4}$ & 7 & $2.7\times10^{-3}$ & 1.1 \\
			\hline
			\bf{B$_\text{max}$} & $1.4\times10^{-7}$ & $2.1\times10^{-8}$ & $1.1\times10^{-2}$ & $1.3\times10^{-4}$ & 82 & $2.2\times10^{-4}$ &  0.1\\ 
			\hline \hline
			\bf{C$_\text{min}$}& $1.7\times10^{-6}$ & $1.9\times10^{-8}$ & $7.6\times10^{-4}$ & $1.3\times10^{-4}$ & 6 & $1.7\times10^{-3}$ & 1.2 \\
			\hline
			\bf{C$_\text{max}$} & $1.4\times10^{-7}$ & $1.9\times10^{-8}$ & $9.3\times10^{-3}$ & $1.3\times10^{-4}$ & 71 & $1.4\times10^{-4}$ &  0.1\\ 
			\hline \hline
			\bf{D$_\text{min}$}& $1.7\times10^{-6}$ & $1.6\times10^{-8}$ & $6.6\times10^{-4}$ & $1.6\times10^{-4}$ & 4 & $2.6\times10^{-4}$ & 1.4 \\
			\hline
			\textcolor{gray}{\bf{D$_\text{max}$} }& $1.4\times10^{-7}$ & $1.6\times10^{-8}$ & $8.0\times10^{-3}$ & $1.6\times10^{-4}$ & 50 & \textcolor{gray}{$2.1\times10^{-5}$} &  0.1\\ 
			\hline \hline
			\bf{E$_\text{min}$}& $1.7\times10^{-6}$ & $2.7\times10^{-8}$ & $1.2\times10^{-3}$ & $3.1\times10^{-4}$ & 4 & $2.5\times10^{-4}$ & 0.8\\
			\hline
			\textcolor{gray}{\bf{E$_\text{max}$}} & $1.4\times10^{-7}$ & $2.7\times10^{-8}$ & $1.4\times10^{-2}$ & $3.1\times10^{-4}$ & 45 & \textcolor{gray}{$2.0\times10^{-5}$} & 0.1 \\ 
			
			\hline
		\end{tabular}  
		\caption {"Extreme" cases of model series A-E: models with the highest possible $g/d$  (and thus $M_\text{gas}(R<90\,\text{au})$) are denoted by the "max", the corresponding lowest models by "min. We give the following parameters: fractional abundance of C$^{18}$O as obtained by matching the observed line profile within the 90\,au snowline radius, the mass of C$^{18}$O within this radius, the H$_2$ mass within the snowline radius, the average $g/d$ within 90\,au, $\alpha_\text{turb}$ and $\alpha_\text{visc}$ for the respective cases. Models~D$_\text{max}$ and E$_\text{max}$ can be excluded as their $\alpha_\text{turb}$ is outside the range we assume.}
		\label{tab:extreme} 
	\end{table*}
\end{center}

\begin{figure}
	\centering
	\includegraphics[width=\columnwidth]{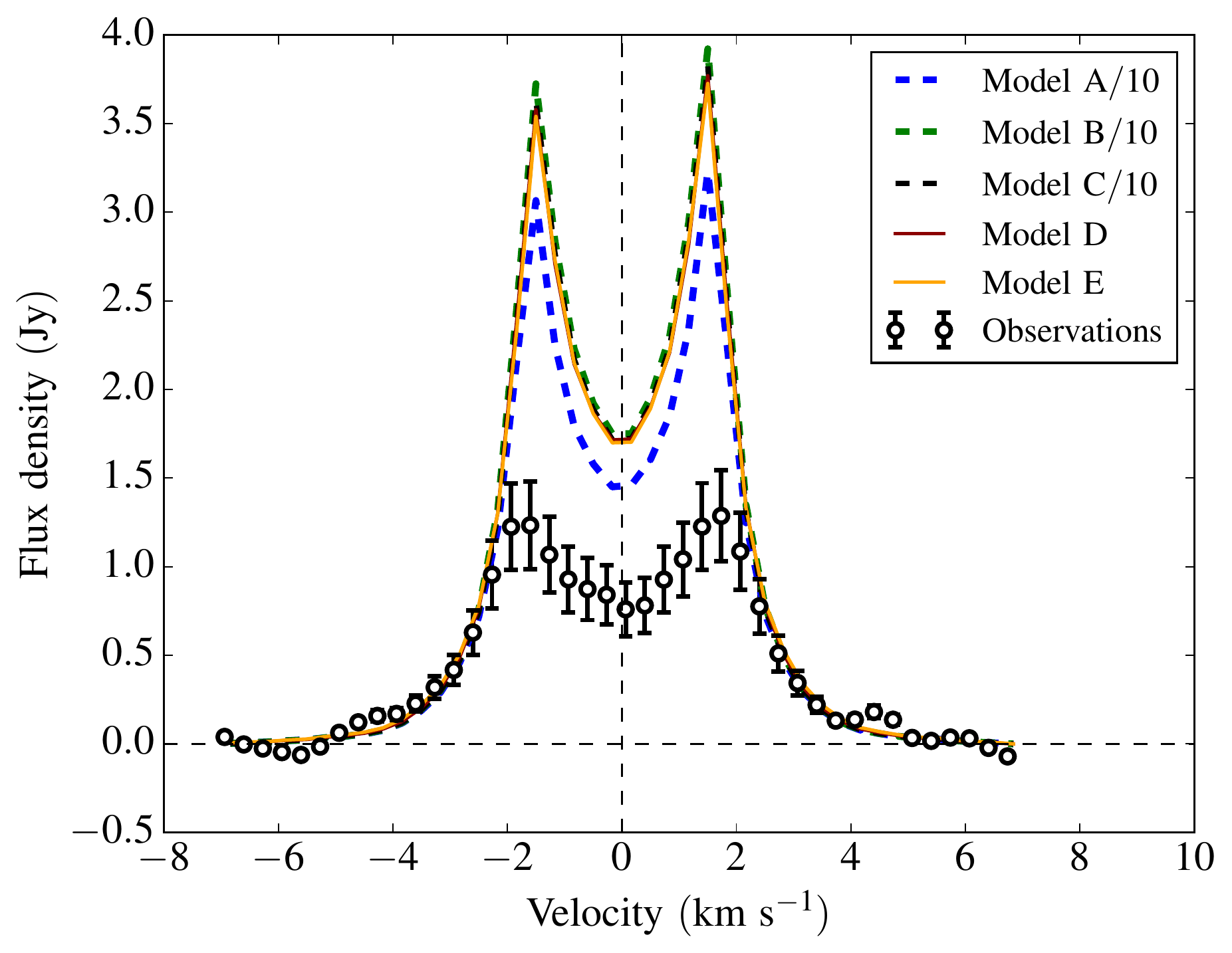}
	\caption{C$^{18}$O line profile for the whole disc for our five fiducial models (colours) and as observed (dots, error bars represent a 10$\%$ flux calibrations uncertainty, \protect\cite{GuidiEtAl2016}). The profiles were centred around a velocity of 0\,km\,s$^{-1}$. All models from the same series (that have C$^{18}$O abundances in the allowed range) have the same line profile (as they have different C$^{18}$O abundances and gas masses, but the same C$^{18}$O mass). Our models match the emission from the inner disc regions (wings of the line profile) very well, whereas they over-predict the emission in the outer disc regions (peaks of the spectrum). Our modelling approach is however best suited for the inner disc regions.}
	\label{fig:wholespectrum}
\end{figure}

\subparagraph{\bf{Modelling the entire disc}}
Finally, for completeness, we compare the synthetic line profile for the whole disc with observations in Figure~\ref{fig:wholespectrum}.
We see that our models over-predict the emission from the outer disc regions, i.e. the emission in the peaks of the spectrum. However, as we mentioned earlier, the SED does not provide information about the structure of the outermost disc regions. Also, we know that there are radial differences in the structure of the outer disc and the inner 240\,au and we therefore limit our attention to the disc inner regions in this paper.
It might be interesting to combine our modelling approach for the inner disc regions with high-resolution imaging of multiple isotopologues in the outer disc regions \protect\citep[see e.g.][]{QiEtAl2011,deGregorioMonsalvoEtAl2013}.

\subsubsection{Physical properties of our models}
\subparagraph{\bf{Gas mass within the snowline radius}}
The match to the observed C$^{18}$O spectrum within the CO snowline which we find for our five models unambiguously constrains the mass of C$^{18}$O in this disc region.
In Equation~\ref{eq:mc18o_90au}, we gave this mass within the snowline radius. We thus calculate the mass of H$_2$ in this disc region from ${M}_{\text{C}^{18}\text{O}}(90\,\text{au})$ by taking into account the abundance of C$^{18}$O given in Table~\ref{tab:90au} and the mass ratio of these molecules ${m_{\text{C}^{18}\text{O}}}/{m_{\text{H}_2}} \approx 14$.
We find that the mass of H$_2$ within the snowline radius is in a range of
${M}_{\text{gas}}(R<90\,\text{au}) \approx (1.3 - 5.0) \times 10^{-3}\,\text{M}_\odot$.

\begin{figure}
	 	\centering
	 	\includegraphics[width=\columnwidth]{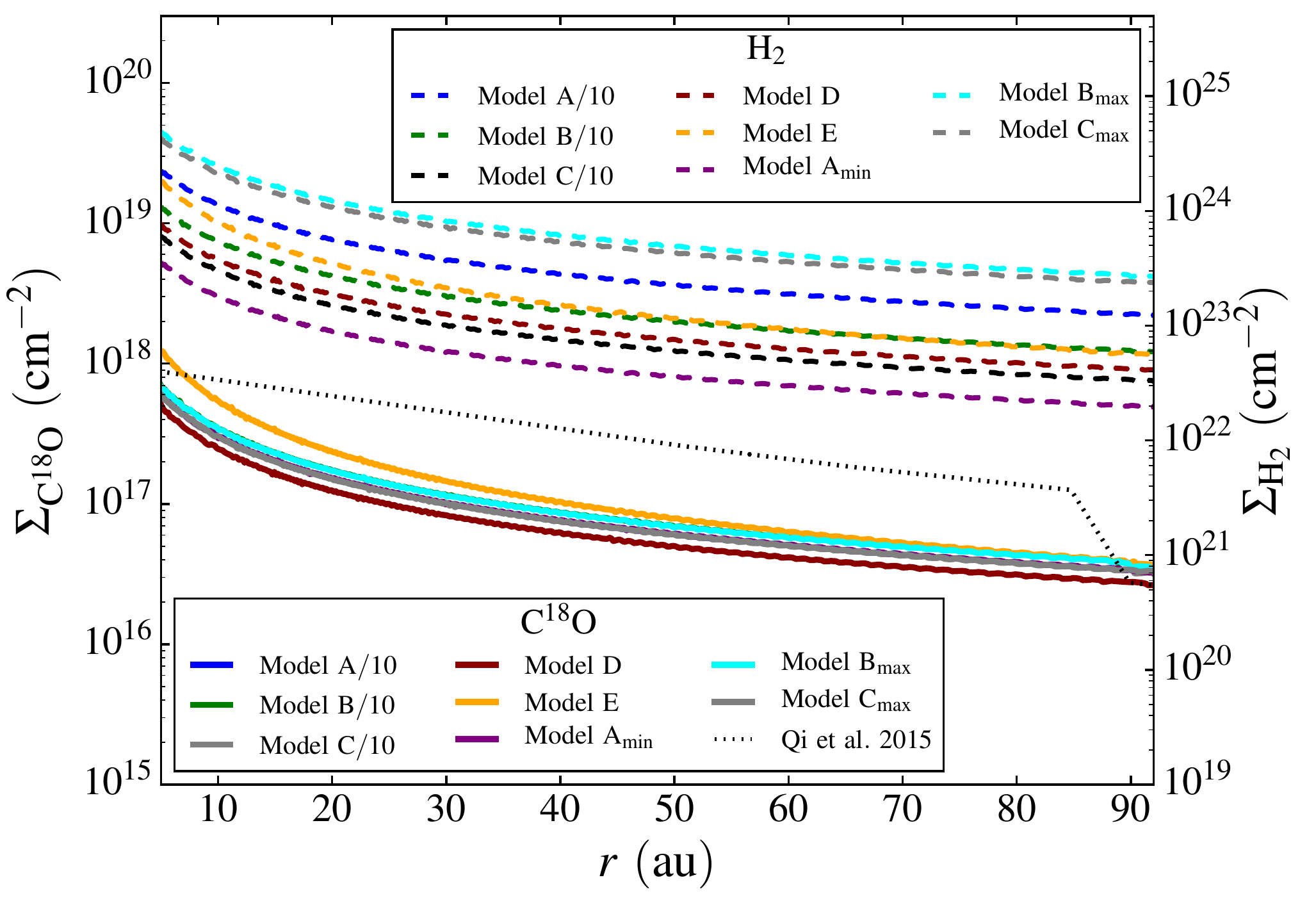}
	 	\caption{Left y-axis: column number density of C$^{18}$O for our models (A-C)/10, D, E,  A$_\text{min}$, B$_\text{max}$ and C$_\text{max}$ for regions within the 90\,au snowline radius, given by the solid colourful lines. We overplot the one obtained by \protect\cite{QiEtAl2015} (dotted black line, extrapolated from 50\,au inwards). Right y-axis: corresponding column density of H$_2$, given by the dashed lines for the individual models. The masses of both C$^{18}$O and H$_2$ that correspond to these column densities are given in Table~\ref{tab:90au}. }
	 	\label{fig:sigma_c18o_h2}
	 \end{figure}

If we add to this the uncertainty in the C$^{18}$O abundance, we obtain the full range of $M_\text{gas}$ that can possibly be present in the disc within 90\,au based on our calculation of the extreme cases (see Table~\ref{tab:extreme}):
\begin{equation}
6.6 \times 10^{-4}\,\text{M}_\odot \lesssim {M}_{\text{gas}}(R<90\,\text{au}) \lesssim 1.1 \times 10^{-2}\,\text{M}_\odot \hspace{2pt}.
\end{equation}
We plot the C$^{18}$O surface number density for our five models and three extreme cases as solid colourful lines in Figure~\ref{fig:sigma_c18o_h2} (left y-axis).
The C$^{18}$O surface density profile derived by \protect\cite{QiEtAl2015} falls within the range shown by our models (black dotted line in Fig.~\ref{fig:sigma_c18o_h2}, extrapolated from 50\,au inwards).
In Figure~\ref{fig:sigma_c18o_h2}, we also give the corresponding H$_2$ column densities in the same plot (dashed lines, right y-axis), where we have employed the C$^{18}$O abundances as listed in Table~\ref{tab:90au}.
	
It is important to note that all eight models yield very similar C$^{18}$O surface densities. However, due to their different C$^{18}$O abundances (see Table~\ref{tab:90au}), their corresponding gas masses within 90\,au are different by approximately an order of magnitude. Not surprisingly, models~B$_\text{max}$ and C$_\text{max}$ yield the highest H$_2$ column densities, given that they have the lowest possible C$^{18}$O abundance and thus the highest gas mass. Model~A$_\text{min}$, on the other hand, has the lowest H$_2$ column density of the models we plot here, as it has the lowest gas mass of all of them. If we were to plot the same for models for (B-D)$_\text{min}$, this would be comparable to A$_\text{min}$.

\subparagraph{\bf{Gas to dust mass ratios}}
Here we present an analysis of the average $g/d$ in the inner disc regions. The $g/d$ values for the individual models are given in Table~\ref{tab:90au} for the fiducial models~(A-C)/10, D and E and for the extreme cases in Table~\ref{tab:extreme}.
It is striking that all models (excluding the extreme cases) have very low $g/d$ values ($9 \lesssim g/d \lesssim 20$). This is significantly lower than the standard  value of 100 as observed in the ISM. However, models~B$_\text{max}$ and C$_\text{max}$ do - by construction - have more ISM-like $g/d$ values ($\sim 80$ and $\sim 70$, respectively). These are the maximum $g/d$ values (and thus the maximum $M_\text{gas}(<90\,\text{au})$) our models can obtain while still matching the C$^{18}$O line profiles, as both of these models have the lowest possible fractional abundance of C$^{18}$O that we consider (see Equation~\ref{eq:minmaxabund}).
We comment more on the possibility of lower C$^{18}$O abundances below.

It is important to note that similarly we can also obtain models based on A-E  - employing the highest possible abundance - that yield the lowest possible g/d while still matching the C$^{18}$O line profile. These are the cases denoted by "min" in Table~\ref{tab:extreme}. For these the $g/d$ values go to values as low as 2 (A$_\text{min}$).

The low value of $g/d\sim55$ that \protect\cite{KamaEtAl2015} infer for the inner disc of HD~163296 using the stellar photosphere is in line with the range of $g/d$ we obtain for the innermost 90\,au.
On the other hand, this range is significantly lower than the value reported by \protect\cite{WilliamsBest2014} ($g/d=170$).
These quantities cannot be compared directly however because, as we pointed out earlier, our results are derived specifically for the $R<90$\,au region, whereas the modelling by \protect\cite{WilliamsBest2014} involves disc emission as a whole and therefore is affected by the assumptions made on the disc vertical structure at large scales inasmuch as this affects the amount of CO that is frozen out. We also note that the very small contribution of such radii to the SED means that the temperature structure of the outer disc is poorly constrained observationally.

\subparagraph{\bf{C$^{18}$O abundance}}

To estimate the maximum gas mass, we adopt the lowest possible value of the abundance of C$^{18}$O (as given in Equation~\ref{eq:minmaxabund}). This corresponds to the lowest CO abundance
measured in the ISM \protect\citep{FrerkingEtAl1982}, combined with the highest isotopic ratio of $^{16}$O to $^{18}$O of 587 \protect\citep{Wilson1999}. This maximum H$_2$ mass sets an upper limit to the possible $g/d$ in our models.

In Section~\ref{subsec:2dstructuremcmax}, we discussed the direct degeneracy between $M_\text{gas}$ and $\alpha_\text{turb}$ in setting the vertical structure of the disc as constrained by the SED. We can see in Table~\ref{tab:extreme} that for some of our models $\alpha_\text{turb}$ could be decreased further, to be compensated with a proportional increase in $M_\text{gas}$, (e.g. models (A-C)$_\text{max}$) if we did not impose a limit on the C$^{18}$O abundance as discussed above. If indeed the C$^{18}$O abundance were a free parameter, our models (A-C) would be compatible with an ISM-like $g/d$ of 100. An assumption of the minimum value for the turbulent mixing strength $\alpha=10^{-4}$ yields a C$^{18}$O abundance as low as $\sim 2.6 \times 10^{-8}$ (in model series~A, corresponding to $g/d \sim 100$). This is lower than the minimum value derived based on the observations of the ISM by a factor of $\sim 5$ . For models~D and E, ISM-like $g/d$ cannot be achieved as we are limited by our lower threshold of $\alpha_\text{turb}$ as discussed above, and therefore the $g/d$ in these models cannot reach higher values than $\sim 10$. We can conclude that $g/d$=100 is possible if one is prepared to assume lower C$^{18}$O abundances. However, this is only true for model series A-C which are the least plausible of our models because their $a_\text{min}$ values of 0.5-0.8\,$\mu$m are only marginally consistent with the result of \protect\cite{GarufiEtAl2014}, where the scattered light observations of HD~163296 imply that the disc surface is dominated by sub-micron-sized grains.

A mechanism to decrease the CO (and isotopologue) abundances, and thus permit a more ISM-like $g/d$, is described in \protect\cite{ReboussinEtAl2015}. Through this mechanism, the C atoms generated through CO photodissociation in the upper layers are effectively removed through formation of species other than CO (e.g. CO$_2$ and CH$_4$). Photodissociation is normally localized in the disc surface, and the C$^{18}$O abundance may be affected only if the CO dissociating photons were able to penetrate to the mid-plane, or if the surface continued to be depleted of CO over very long time-scales. Thus far, comparison of the CO (and isotopologue) abundance to the H$_2$ density has only been possible for one, particularly old disc, TW~Hya \protect\citep{FavreEtAl2013,SchwarzEtAl2016}. These works measure the abundance of CO and its isotopologues to be about 100 times lower than their ISM values.

\subparagraph{\bf{Dust dynamics}}
The outer regions of the disc are known to be deficient in sub-mm grains (as deduced
from the outer radius in the sub-mm continuum compared with that in
CO; see also \protect\cite{GuidiEtAl2016}). Such concentration of sub-mm dust
in the inner disc can be explained in terms of drag-mediated migration
of solids. We have used the mid-plane density and temperature profiles of our favoured models to estimate the Stokes number (ratio of drag time-scale to dynamical time-scale)
as a function of grain size. We find a Stokes number of close to unity (which corresponds to maximal radial migration) for mm-sized grains in the region of the CO snowline.

The majority of our models have $g/d$ ratios that are considerably below the ISM ratio of 100. Indeed, we can only approach this value if we assume a very low fractional abundance of C$^{18}$O {\it and} assume a grain population that is highly depleted in sub-micron grains (this latter is required in order not to over-predict the infrared flux, given the relatively high temperatures obtained in the case of dust supported in gas-rich discs).
However, \protect\cite{GarufiEtAl2014} find from their studies of scattered light that there are sub-micron-sized particles present in the inner disc regions, thus the $a_\text{min}=0.8$ and 0.5\,$\mu$m in models~B$_\text{max}$ and C$_\text{max}$ are only marginally consistent with this requirement. We conclude that the available data require significant deviation from primordial conditions, either in terms of depletion of gas or else in terms of depletion of small grains.

\subparagraph{\bf{Mid-plane gas-to-dust ratios}}
 
 \begin{figure}
 	\centering
 	\includegraphics[width=\columnwidth]{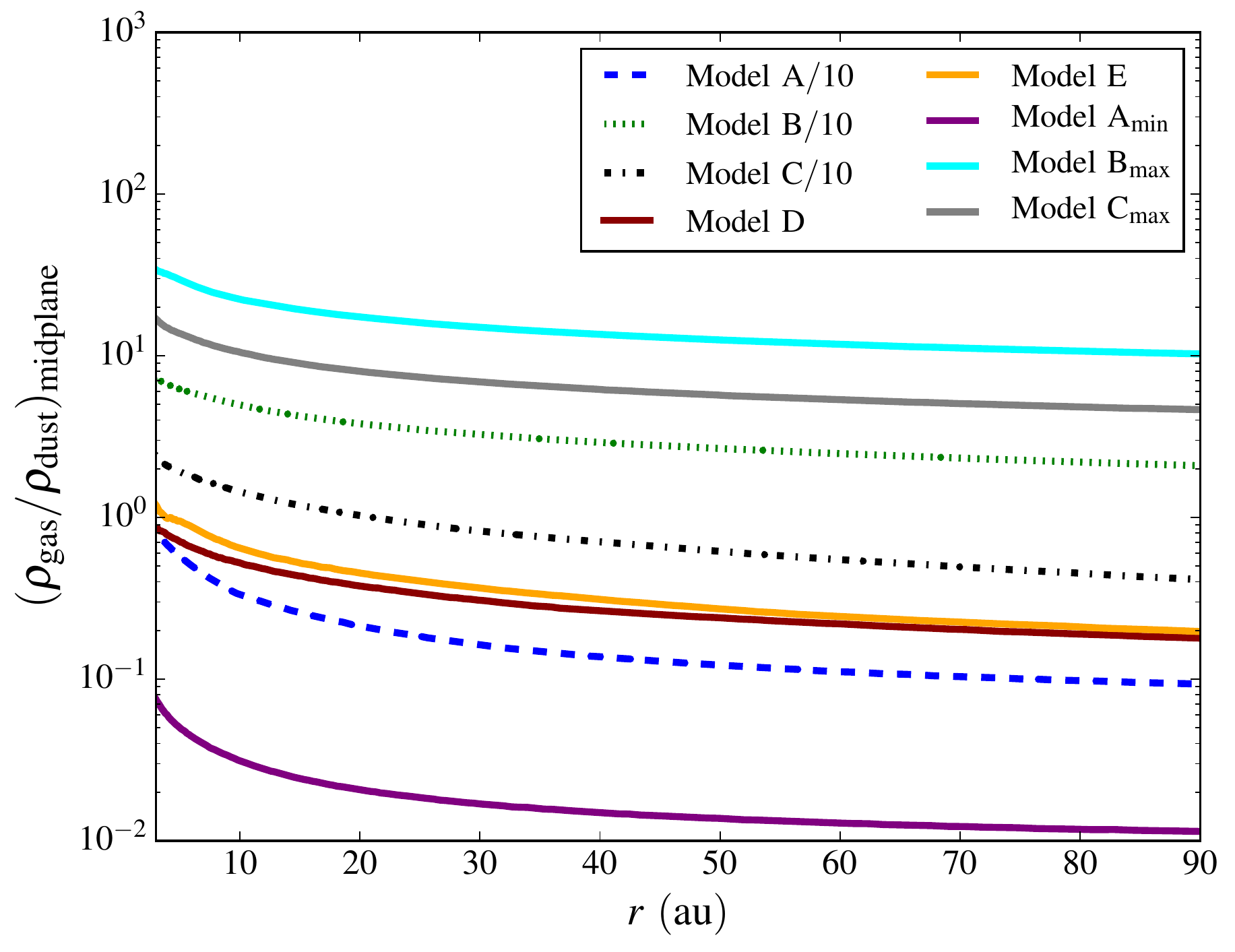}
 	\caption{Gas-to-dust mass ratio in the disc mid-plane for our models (A-C)/10, D, E,  A$_\text{min}$, B$_\text{max}$ and C$_\text{max}$ in the inner disc regions ($R<90$\,au).}
 	\label{fig:dust-gas-mid}
 \end{figure}
It is important to note that the $g/d$ values we have presented so far are average values.
Focusing on the ratio of the gas-to-dust density in the inner disc mid-plane now, we show a plot of their ratio in Figure~\ref{fig:dust-gas-mid}. All our models show very low ratios of $\rho_\text{gas}/\rho_\text{dust}$ in the mid-plane (between $\sim 0.01 - 20$) within a radius of 90\,au. It is interesting to mention that model~B/10, B$_\text{max}$ and C$_\text{max}$ have the highest ratio (around $\sim 10$). Model  A$_\text{min}$ has a $g/d$ of as low as $\sim 0.01$, which is not surprising given that its average $g/d$ ratio is $\sim 2$ and thus the lowest possible in all our models. The fact that $\rho_\text{gas}/\rho_\text{dust}$ in the mid-plane is lower than the average $g/d$ as discussed above is a result of the dust settling in our models.

\subparagraph{\bf{Viscosity}}
Using the masses of H$_2$ within a radius of 90\,au (Table~\ref{tab:90au}), we calculate the viscosity parameter $\alpha_\text{visc}$ of the inner disc regions. The mass accretion rate of HD~163296 is measured to be within the range (0.8-4.5)$\times 10^{-7}$\,M$_\odot$ yr$^{-1}$ \protect\citep{GarciaLopezEtAl2015}. 
The viscosity parameter is given by 
\begin{equation}
\alpha_\text{visc}=\left(\frac{H}{r}\right)^{-2} \frac{\tau_\text{dyn}}{\tau_\text{visc}} \hspace{2pt}.
\label{eq:alphavisc}
\end{equation}
The dynamical time-scale of our models at 90\,au is 
\begin{equation}
\tau_\text{dyn}(90\,\text{au}) = \Omega_\text{k}^{-1} \sim 100\,\text{yr} \hspace{2pt}.
\end{equation}
The time-scale of the flow can be obtained by
\begin{equation}
\tau_\text{flow}(90\,\text{au}) =\frac{M_\text{gas}(90\,\text{au})}{\dot{M}} \hspace{2pt},
\end{equation}
which will thus vary depending on the masses of H$_2$ within 90\,au as given in Table~\ref{tab:90au}. The scale-height of our models is $H/r \sim 0.1$. Equating $\tau_\text{visc}$ and $\tau_\text{flow}$, we obtain from Equation~\ref{eq:alphavisc} a range of $\alpha_\text{visc}$ of 
\begin{equation}
    0.2 \times \left(\frac{\dot{M}}{10^{-7}\text{M}_{\odot}\textrm{yr}^{-1}}\right) < \alpha_\text{visc} <  0.7 \times \left(\frac{\dot{M}}{10^{-7}\text{M}_{\odot}\textrm{yr}^{-1}}\right) \hspace{2pt},
\end{equation}
as given in detail for the respective models in Table~\ref{tab:90au}.
When considering the models with minimum and maximum $M_\text{gas}$ (Table~\ref{tab:extreme}), the range of $\alpha_\text{visc}$ we obtain is between 0.1 and 1.4.
The values we obtain for $\alpha_\text{visc}$ are much higher than those found in magneto-hydrodynamical simulations of the outer regions of discs in T Tauri stars \protect\citep{SimonEtAl2013a}, suggesting that more efficient angular momentum transport (such as that linked to a large scale net magnetic field and associated wind; \protect\cite{SimonEtAl2013b}) may be required. 
We also note that our derived $\alpha_\text{visc}$ values are around two orders of magnitude greater than the maximum values of $\alpha_\text{turb}$ allowed by our modelling. This suggests that the efficient transport of angular momentum in this disc is not accompanied by the vigorous level of vertical motions that would be expected in the case of a turbulent viscosity model.

\begin{figure}
 	\centering
 	\includegraphics[width=\columnwidth]{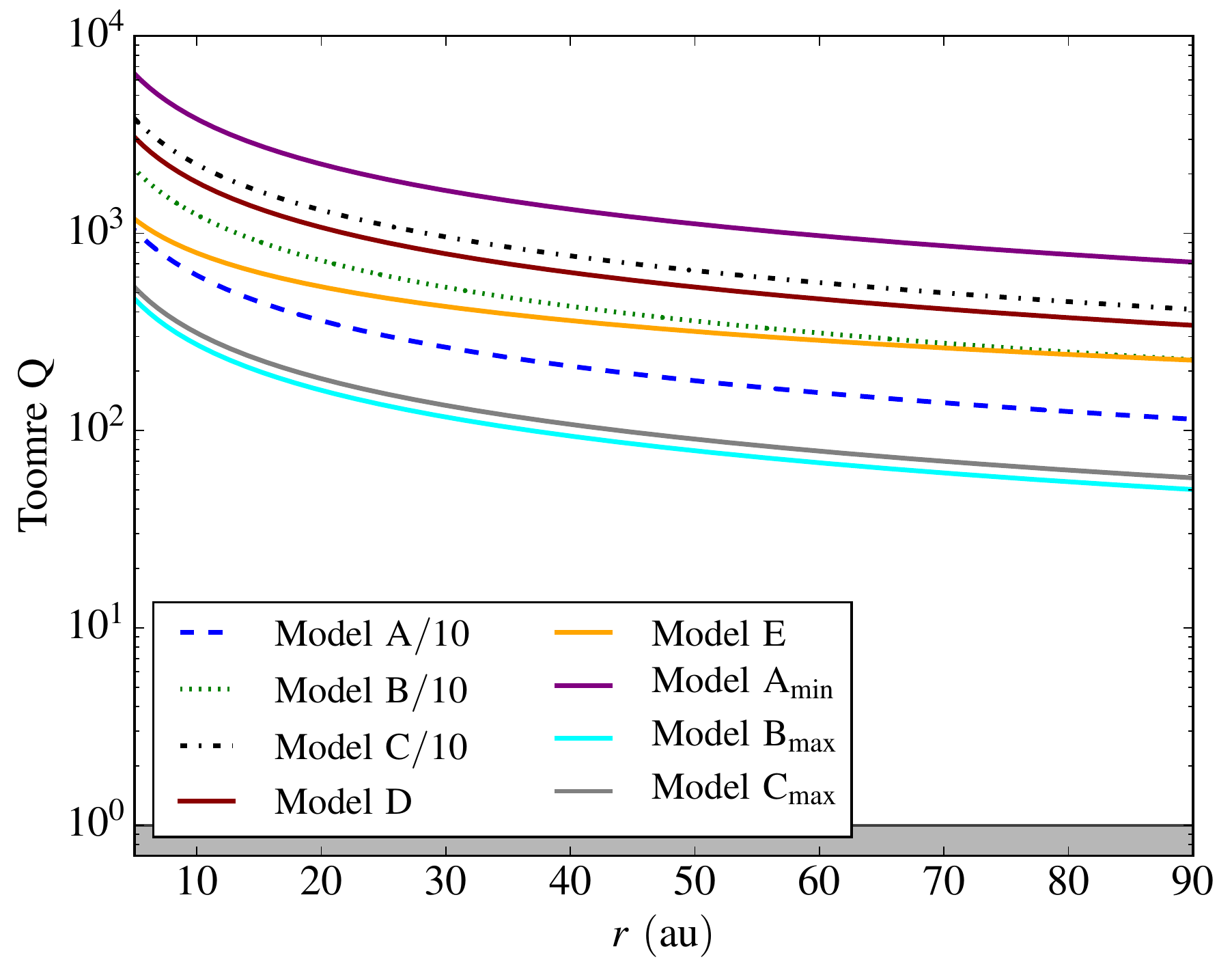}
 	\caption{Range of Toomre Q parameter for our models (A-C)/10, D, E,  A$_\text{min}$, B$_\text{max}$ and C$_\text{max}$ which we had found to match the observed SED, the snowline radius and the C$^{18}$O emission within 90\,au. The region where $Q<1$ (and the disc potentially gravitationally unstable) is shaded in grey. We find that none of our models reaches this critical regime and all are well above the threshold.}
 	\label{fig:toomre}
 \end{figure}

\subparagraph{\bf{Gravitational instability}}
We can estimate the gravitational stability of HD~169392 by evaluating the Toomre stability parameter
\begin{equation}
Q=\frac{c_\text{s}\Omega}{\pi \Sigma G}<1
\end{equation}
in our models, where $c_\text{s}$ is the sound speed, $\Omega$ the Keplerian frequency and $\Sigma$ the disc surface density \protect\citep{Toomre1964} and where self-gravity becomes important at a $Q$ value of close to unity. 
We thus study the mid-plane $Q$ parameter as a function of radius in all our eight disc models ((A-C)/10, D, E,  A$_\text{min}$, B$_\text{max}$ and C$_\text{max}$) and plot the results in Figure~\ref{fig:toomre}.
We find that all of our models are well above the threshold value of $Q=1$.
This implies that none of the models are close to being gravitationally unstable at radii $<90$\,au; we, however, caution that we cannot assess this quantity in the outer disc, given the sensitivity to the degree of freeze-out in these outer disc regions.

\section{Conclusions}
\label{sec:summaryconclusions}
We combine SED fitting, the location of the CO snowline and spatially resolved C$^{18}$O line emission to help resolve degeneracies in the determination of protoplanetary disc properties, using the example of HD~163296, of which we estimate the properties. We draw the following main conclusions from this work:
\begin{enumerate}
	\item Any one of the aforementioned diagnostics is on its own insufficient to robustly determine the disc properties; however, we demonstrate that together they become much more powerful tools. SED and CO snowline fitting alone could result in a disc mass almost an order of magnitude higher than the mass obtained when C$^{18}$O observations are included.
	\item The observed C$^{18}$O line flux, together with SED and CO snowline
	modelling, unambiguously indicates the mass of gas-phase C$^{18}$O within the 90\,au snowline radius, ${M}_{\text{C}^{18}\text{O}}(R<90\,\text{au})\approx (2 - 3) \times 10^{-8}$\,M$_\odot$.
	We obtain a total gas mass ${M}_{\text{gas}}(R<90\,\text{au})\approx (0.7 - 11) \times 10^{-3}$\,M$_\odot$ within the snowline radius, taking into account the uncertainties in the fractional abundance of C$^{18}$O. 

	\item Our modelling approach is best suited for the inner disc regions (within the snowline radius). The emission from the outer disc regions is crucially dependent on the vertical temperature structure and the location of the CO ice surface, so we do not aim to match these. 
	 From this, we can conclude that it is important to constrain the vertical temperature of the disc well through physically consistent SED models for the inner disc (as we presented here) and combine these with, for example, high-resolution imaging of multiple CO isotopes in the outer disc \protect\citep[see e.g.][]{QiEtAl2015}.
	\item For the range of $\alpha_\text{turb}$ from $(0.1-6.3) \times 10^{-3}$, most of our models of HD~163296 imply gas-to-dust mass ratios in the range $g/d=10-20$, significantly lower than the ISM value of 100. If we are prepared to also consider models with minimum dust grain sizes of $\sim 0.5\,\mu$m that are not fully consistent with scattered light observations \protect\citep{GarufiEtAl2014} that also have very low (high) fractional abundance of C$^{18}$O, models with $g/d$ as large as 80 (as small as 2) also match the observations. On top of this and only for these extreme models, $g/d=100$ may be achieved if the CO abundance is anomalous due to e.g. C-sequestration.
	\item We obtain a high $\alpha_\text{visc} \sim 0.2 - 0.7$ for our models of the inner disc regions, or even up to values of $\alpha_\text{visc} \sim 1.4$ (0.1), if we allow for C$^{18}$O to be very over- (under-)abundant with respect to the ISM abundances. 
	The notably high ratio of $\alpha_\text{visc}$ to $\alpha_\text{turb}$ provides evidence against a turbulent model for angular momentum transport in this disc.
	\item From analysis of the temperature and density profiles obtained from our models, we find that the disc is not likely to be susceptible to gravitational instability. 
\end{enumerate}
The approach to interpretation outlined in this paper will allow us to maximize the value of existing and future high-quality observations with ALMA. This work stresses the importance of C$^{18}$O observations especially for the warm Herbig Ae discs, which are the prime targets for the application of the methods outlined in this paper.

\section*{Acknowledgements}
We would like to thank the anonymous reviewer for their thorough report and suggestions that have improved the work. This paper makes use of the following ALMA data: ADS/JAO.ALMA$\#$2011.0.00010.SV. ALMA is a partnership of ESO (representing its member states), NSF (USA), and NINS (Japan), together with NRC (Canada) and NSC and ASIAA (Taiwan), in cooperation with the Republic of Chile. The Joint ALMA Observatory is operated by ESO, AUI/NRAO and NAOJ.

This work has been supported by the DISCSIM project, grant agreement 341137 funded by the European Research Council under ERC-2013-ADG. DMB is funded by this ERC grant and an STFC studentship. OP is supported by the Royal Society Dorothy Hodgkin Fellowship. During a part of this project OP was supported by the European Union through ERC grant number 279973. TJH is funded by the STFC consolidated grant ST/K000985/1. We thank the DISCSIM group at the IoA Cambridge and especially Richard Booth and Attila Juh{\'a}sz for helpful discussions. 




\bibliographystyle{mnras}
\bibliography{bibliography.bib} 








\bsp	
\label{lastpage}
\end{document}